\documentclass[twocolumn]{aastex631}

\usepackage{epsfig}
\usepackage{xcolor}

\usepackage{graphicx}
\usepackage{float}
\usepackage[caption =false]{subfig}
\usepackage{graphicx}
\usepackage{multirow}
\usepackage{enumitem,kantlipsum}
\newcommand{\megasaura}{M\textsc{eg}a\textsc{S}a\textsc{ura}}
\newcommand{\megasauralong}{The Magellan Evolution of Galaxies Spectroscopic and Ultraviolet Reference Atlas}

\newcommand{\HST}{{\it HST}}
\newcommand{\kms}{km s$^{-1}$}

\shorttitle{Ionized gas outflow from a LyC emitter}
\shortauthors{Mainali et al.}

\begin{document}

\title{The connection between galactic outflows and the escape of ionizing photons}

\author[0000-0003-0094-6827]{Ramesh Mainali}
\affiliation{Observational Cosmology Lab, Code 665, NASA Goddard Space Flight Center, Greenbelt, MD 20771, USA}
\affiliation{Department of Physics, The Catholic University of America, Washington, DC 20064, USA}

\author[0000-0002-7627-6551]{Jane R. Rigby}
\affiliation{Observational Cosmology Lab, Code 665, NASA Goddard Space Flight Center, Greenbelt, MD 20771, USA}

\author[0000-0002-0302-2577]{John Chisholm}
\affiliation{Astronomy Department, University of Texas at Austin, 2515 Speedway, Stop C1400, Austin, TX 78712-1205, USA}

\author[0000-0003-1074-4807]{Matthew Bayliss}
\affiliation{Department of Physics, University of Cincinnati, Cincinnati, OH 45221, USA}

\author[0000-0002-3120-7173]{Rongmon Bordoloi}
\affiliation{Department of Physics, North Carolina State University, Raleigh, North Carolina, 27695}

\author[0000-0003-1370-5010]{Michael D. Gladders}
\affiliation{Department of Astronomy \& Astrophysics, The University of Chicago, 5640 S.~Ellis Avenue, Chicago, IL 60637, USA}

\author[0000-0002-9204-3256]{T. Emil Rivera-Thorsen}
\affiliation{The Oskar Klein Centre, Department of Astronomy, Stockholm University, AlbaNova, SE-10691 Stockholm, Sweden}

\author[0000-0003-2200-5606]{H\r{a}kon Dahle}
\affiliation{Institute of Theoretical Astrophysics, University of Oslo, P.O. Box 1029, Blindern, NO-0315 Oslo, Norway}

\author[0000-0002-7559-0864]{Keren Sharon}
\affiliation{University of Michigan, Department of Astronomy, 1085 South University Avenue, Ann Arbor, MI 48109, USA}

\author[0000-0001-5097-6755]{Michael Florian}
\affiliation{Steward Observatory, University of Arizona, 933 North Cherry Ave., Tucson, AZ 85721, USA}

\author[0000-0002-4153-053X]{Danielle A. Berg}
\affiliation{Astronomy Department, University of Texas at Austin, 2515 Speedway, Stop C1400, Austin, TX 78712-1205, USA}

\author{Soniya Sharma}
\affiliation{Observational Cosmology Lab, Code 665, NASA Goddard Space Flight Center, Greenbelt, MD 20771, USA}

\author[0000-0002-2862-307X]{M. Riley Owens}
\affiliation{Department of Physics, University of Cincinnati, Cincinnati, OH 45221, USA}

\author{Karin Kjellgren}
\affiliation{The Oskar Klein Centre, Department of Astronomy, Stockholm University, AlbaNova, SE-10691 Stockholm, Sweden}

\author[0000-0001-6505-0293]{Keunho J. Kim}
\affiliation{Department of Physics, University of Cincinnati, Cincinnati, OH 45221, USA}

\author{Julia Wayne}
\affiliation{Department of Physics, University of Cincinnati, Cincinnati, OH 45221, USA}

\email{ramesh.mainali@nasa.gov, jane.r.rigby@nasa.gov}









\begin{abstract}

We analyze spectra of a gravitationally lensed galaxy, known as the Sunburst Arc, that is leaking ionizing photons, also known as the Lyman continuum (LyC).  Magnification from gravitational lensing permits the galaxy to be spatially resolved into one region that leaks ionizing photons, and several that do not. Rest-frame ultraviolet and optical spectra from Magellan target ten different regions along the lensed Arc, including six multiple images of the LyC leaking region, as well as four regions that do not show LyC emission. The rest-frame optical spectra of the ionizing photon emitting regions reveal a blue-shifted ($\Delta V$=27 km s$^{-1}$) broad emission component (FWHM=327 km s$^{-1}$) comprising 55\% of the total [OIII] line flux,  in addition to a narrow component (FWHM = 112 km s$^{-1}$), suggesting the presence of strong highly ionized gas outflows. This is consistent with the high-velocity ionized outflow inferred from the rest-frame UV spectra. In contrast, the broad emission component is less prominent in the non-leaking regions, comprising $\sim$26\% of total [OIII] line flux. The high ionization absorption lines are prominent in both leaker and non-leaker but low ionization absorption lines are very weak in the leaker, suggesting that the line of sight gas is highly ionized in the leaker. Analyses of stellar wind features reveal that the stellar population of the LyC leaking regions is considerably younger ($\sim$3 Myr) than the non-leaking regions ($\sim$12 Myr), highlighting that stellar feedback from young stars may play an important role in ionizing photon escape.

\end{abstract}



\section{Introduction} \label{sec:intro}

Understanding the epoch of reionization, when the last phase transition of the Universe occurred, is one of the longstanding goals of extragalactic astronomy. The massive stars residing in these high-redshift star-forming galaxies are suspected to be the dominant ionizing agents that drive reionization \citep[e.g.][]{Robertson2015,Finkelstein2019}, although some studies suggest that low luminosity active galactic nuclei (AGN) may play a significant role in reionization \citep[e.g.][]{Madau2015,Kulkarni2019}. The study of reionization requires understanding the rate at which ionizing photons are injected into the intergalactic medium (IGM) by the galaxy population, known as ionizing emissivity \citep[e.g.][]{Robertson2015, Finkelstein2019}. In practice, this quantity can be measured by combining three key factors: the rest-frame ultraviolet (UV) luminosity density ($\rm \rho_{UV}$), the ionizing photon production efficiency ({$\rm \xi_{ion}$}), and the fraction of ionizing photons that escapes from galaxies ($\rm f_{esc}$). Soon {\it The James Webb Space Telescope} ({\it JWST}) will measure the first two quantities at the epoch of reionization, however, the measurement of $\rm f_{esc}$ will heavily rely on indirect probes at high redshift or work on analogs at lower redshift since ionizing fluxes cannot be measured at $z>6$ due to the large opacity of the IGM. Further progress hinges on understanding the detailed astrophysics of ionizing photons escape from Lyman continuum (LyC) emitting systems at lower redshift.

One of the main suspects in aiding escape of ionizing photons from galaxy is galaxy scale outflows \citep[e.g.][]{Weiner2009,Heckman2011,Newman2012, Jones2013,Bordoloi2014,Rubin2014, Alexandroff2015, Chisholm2017,Kim2020}. These outflows are expected to remove surrounding neutral gas, thus clearing a pathway for ionizing photons to efficiently escape. However, it is often challenging to reconcile the timescales of such strong galactic outflows with the timescale of production and escape of ionizing photons in a galaxy. This has been demonstrated in cosmological simulations where a detailed treatment of the multiphase interstellar medium (ISM) and feedback were possible, yet very low $\rm f_{esc}$ of only a few percent level were predicted \citep[e.g.][]{Gnedin2008,Kim2013,Wise2014, Ma2015,Rosdahl2018}. In these simulations, most of the ionizing photons are consumed by surrounding neutral gas clouds \citep[e.g.][]{Ma2015, Kimm2017, Kakiichi2021}. The feedback from supernovae activity (SNe) may help in the ionizing photon escape but SNe first occur after about 3 Myr of massive star formation and peaks around 10 Myr. On the other hand, the ionizing photon production from massive stars begins to decline after 3 Myr, and thus allows for a very narrow timing window for the escape of ionizing photons \citep{Ma2015}.

\begin{figure*}
\centering

\includegraphics[scale=0.55]{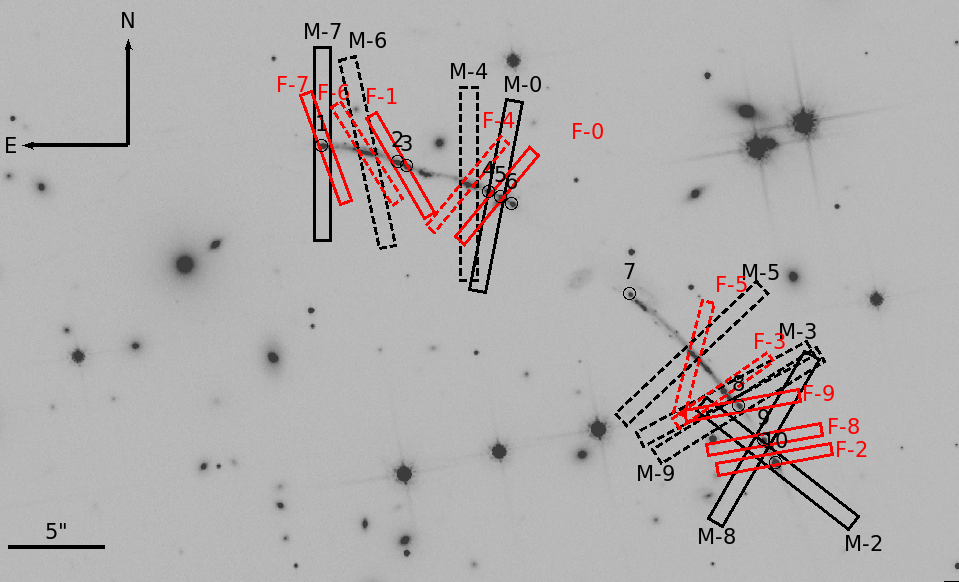}

\caption{HST F814W image of the Sunburst Arc showing the positions of FIRE and MagE slits. The red rectangles show the position and orientation of 10 FIRE slits, and the black rectangles show the position and orientation of 9 MagE slits. The solid and dashed rectangles indicate LyC photons leaking and non-leaking knots, respectively. The numbers indicate multiple images of a LyC emitting region.}
\label{fig:finder}
\end{figure*}

Recent progress on the above picture has come from high-resolution cosmological simulations of LyC emitting sources in the EoR \citep{Ma2020}, studying $z > 5$ simulated galaxies from the Feedback in Realistic Environments project (FIRE: \citealt{Hopkins2018}). The latest version of the simulations incorporates the multi-phase
ISM, star formation, and stellar feedback. The study finds that a majority of LyC escape comes from the very young ($<$10 Myr), vigorously star-forming regions of a galaxy, with negligible contribution from an older ($>10$ Myr) stellar population. The LyC escaping sites are often characterized by feedback-driven kpc-scale superbubbles which clear out the neutral gas column to allow the escape of ionizing photons.

In order to test the theoretical picture discussed above, we must study LyC emitting galaxies in detail to determine the astrophysics that regulate the escape of LyC photons into the IGM. Galaxies at redshifts above $\sim 0.1$ are in general too distant to study in any detail. Rest-UV spectra of nearby galaxies require observations with the Cosmic Origins Spectrograph on HST, whose 2\farcs5 aperture makes it impossible to pinpoint LyC sources on scales smaller than a 100s of pc for the closest galaxies. However, the lower redshift LyC emitters  are very rare; only 3 have been discovered at $z < 0.1$ \citep{Leitet2011, Leitet2013, Leitherer2016}. The average measured LyC escape fractions are found to be only a few percent suggesting very little ionizing radiation of stellar origin escapes from galaxies. At slightly higher redshifts, $z \sim$ 0.1--0.4, ``Green Pea''
galaxies are found to be LyC emitters, with escape fractions up to $\sim 75\%$
\citep{Borthakur2014, Izotov2016, Izotov2018, Izotov2021,Flury2022a}. 
Recently, several programs have discovered LyC emitting galaxies at $1.5 \lesssim z \lesssim 4$; only a handful of these galaxies show large LyC escape fractions \citep{Vanzella2016, Shapley2016, Bian2017, Vanzella2018, Fletcher2019, Steidel2018, Rivera-Thorsen2019}. Overall, the results from lower redshift LyC studies suggest that the LyC escape from an average galaxy population is significantly lower than that required to drive cosmic reionization. It is thus crucial to understand the key differences between lower redshift and reionization era galaxies, and how they impact the overall escape of LyC photons.

Highly magnified lensed galaxies at $z>1$ with significant LyC leakage may present the most tractable sites to study the escape of ionizing
photons at any redshift. This point is powerfully demonstrated by the 
\emph{Sunburst Arc} at z = 2.37 \citep{Dahle2016, Rivera-Thorsen2017, Rivera-Thorsen2019, Chisholm2019}, also known as PSZ1-ARC G311.6602–18.4624, one of the brightest gravitationally lensed galaxies known. {\it HST} observations in the rest-frame LyC (the observed F275W band) clearly show that this galaxy appears 12 times in the image plane with a significant amount of ionizing photons leakage \citep{Rivera-Thorsen2019}. The LyC emitting region has a highly unusual, triple-peaked  Lyman-$\alpha$ line profile, which \citet{Rivera-Thorsen2017} interpret as evidence that the radiation escapes through a narrow, empty channel in an otherwise optically thick medium.

This paper leverages rest-frame UV and optical spectroscopy of multiple distinct regions within the Sunburst arc, to compare regions that are leaking LyC photons to those that do not. This allows us to connect the properties of the outflows to their local driving sources. Rest-frame ultraviolet spectra from the Magellan Echellette \citep[MagE:][]{Marshall2008} instrument on the Magellan/Baade telescope constrain the massive stellar population as well as the interstellar medium.  Rest-frame optical spectra from the Folded-port InfraRed Echelle \citep[FIRE:][]{Simcoe2008} instrument on Magellan/Baade constrain the nebular gas. 

The paper is organized as follows.  We discuss our observational strategy and data reduction in \S2.   In \S3, we discuss the methodology, and present analysis in \S4. We discuss implications for the physical picture of ionizing photon escape and inferred size scale of outflowing gas in \S5, and summarize our findings in \S6.
Throughout this paper, we adopt a $\Lambda$-dominated, flat universe
with $\Omega_{\Lambda}=0.7$, $\Omega_{M}=0.3$ and
$\rm{H_{0}}=70\,\rm{h_{70}}~{\rm km\,s}^{-1}\,{\rm Mpc}^{-1}$. All
magnitudes are quoted in the AB system. Equivalent widths are quoted in rest-frame, unless stated otherwise. 

\begin{table*}
\hspace{-2 cm}
\begin{tabular}{lccccccccc}
\hline
FIRE slit & MagE slit	& RA		&	DEC	&	UT Date	 & time (ks) & LyC leaker  & z$_{spec}$ & PA  & Magnification($\mu$)\\ \hline
F-0	& M-0 &	15:50:04.317	&	-78:10:59.87	&	2016 Mar 30		&	3.6		&	Yes	& 2.37014 & 140.2	& 7.8$^{+15}_{-5.9}$\\
F-1	& \ldots	& 15:50:06.008	&	-78:10:58.07	&	2017 Mar 28		&	3.6		&	Yes	& 2.37009 & 30.2 & 40.3$^{+1.0}_{-4.7}$\\	
F-2	& M-2 &	15:49:59.662	&	-78:11:13.48	&	2017 Mar	28		&	3.6		&	Yes	& 2.37017 & 100.2 & 30.7$^{+6.1}_{-2.9}$ \\
F-3	& M-3 &	15:50:00.488 	&	-78:11:10.00	&	2017 Mar	28		&	4.8		&	No	& 2.37025 & 125.2 & 36.7$^{+4.5}_{-4.9}$\\
F-4	& M-4 &	15:50:04.824 	&	-78:10:59.20	&	2017 Mar	29		&	3.6		&	No	& 2.37073 & 140.2 & 14.7$^{+0.7}_{-2.6}$\\
F-5	& M-5 &	15:50:01.005	&	-78:11:08.07	&	2017 Mar	29		&	2.7		&	No	& 2.37086 & 165.2 & 49.4$^{+5.0}_{-11.3}$\\
F-6	& M-6 &	15:50:06.557	&	-78:10:57.52	&	2017 Mar	29		&	2.7		&	No	& 2.37021 & 32.2 & 141.0$^{+35.0}_{-27.0}$\\
F-7	& M-7 &	15:50:07.238	&	-78:10:57.22	&	2017 Mar	29		&	2.7		&	Yes	& 2.37044 & 20.2 & 34.4$^{+2.4}_{-5.6}$\\
F-8	& M-8 &	15:49:59.834	&	-78:11:12.48	&	2017 Aug	28		&	3.6		&	Yes	& 2.37024 & 100.2 & 28.9$^{+6.1}_{-2.9}$\\
F-9	& M-9 &	15:50:00.251	&	-78:11:10.71	&	2017 Aug	27		&	5.4		&	Yes	& 2.37030 & 100.2 & 30.9$^{+3.8}_{-3.2}$\\
\hline
\end{tabular}
\caption{Observation log for Magellan/FIRE.  The first and second columns represent FIRE slit names and the closest corresponding MagE slit names.  The  remaining columns are Right ascension (J2000), Declination (J2000), Universal date, Universal time, integration time in kiloseconds, whether that source is a LyC leaker according to the \HST\ F275W imaging, spectroscopic redshift measured from optical lines, the position angle (degrees E of N) of the FIRE slit  and the average magnification factor ($\mu$) within the slit.}
\label{table:observation}
\end{table*}

\section{Observations}

\subsection{Magellan/FIRE near-IR spectroscopy}

We observed the Sunburst Arc over the course of three observing runs from 2016--2018. We used the FIRE \citep{Simcoe2008} instrument on Magellan in echelle mode, providing continuous spectral coverage for wavelengths between 0.82 and 2.51 $\rm \mu$m. We adopted a slit width of 0\farcs6, resulting in a resolving power of $R = 6000$. Ten different regions within the arc are observed, including both LyC-emitting and non-emitting regions. The orientations of the 10 different slits are shown in Figure~\ref{fig:finder}. The exposures were carried out using two dither positions separated by 3\farcs0. We obtained total exposure of 9.08 hr with an average exposure of 54.5 min per slit position. Slits were positioned on the sky by directly acquiring one of two nearby reference stars (see Figure~\ref{fig:finder}) and then applying small ($\lesssim$11\arcsec) coordinated offsets to move the slit onto the desired location along the arc.

The FIRE spectra were reduced using standard routines in the FIREHOSE data reduction pipeline\footnote{wikis.mit.edu/confluence/display/FIRE/FIRE+Data+Reduction}. The pipeline calculates two-dimensional sky models iteratively, following \citet{Kelson2003}. The wavelength solutions are provided by fitting OH skylines in the science spectra. Flux calibration and telluric corrections are applied using  A0V star observations that were obtained immediately before or after science observations. The one-dimensional spectra were extracted by using optimal extraction, obtained by calculating spatial profiles of the strongest, well-isolated emission lines in the FIRE spectra (i.e. [OIII]$\lambda$5007). 

\subsection{Near-infrared HST grism spectroscopy}
Near-infrared spectra of the Sunburst Arc were obtained using the HST WFC3/IR G141 grism from program GO-15101 (PI: H. Dahle). The WFC3/IR G141 grism covers wavelengths from 1.075 $\mu m$ to 1.7 $\mu m$ and has a resolving power of $R \sim 130$.  At the redshift of Sunburst Arc ($z=2.37$), the G141 grism covers rest-frame wavelength of $\sim$3200 to 5040~\AA, allowing measurements of all emission lines from [O~II]~3727 to [O~III]~5007. The grism observations were performed at two different orientation angles (27.37$^{\circ}$ and 355.37$^{\circ}$), as is standard practice to mitigate contamination from cluster galaxies.  The total grism integration time was 8.42 Ks.

The HST grism spectra were reduced following the procedures described in \citet{Florian2021}.
To summarize, we used the software package Grizli13 \citep{Grizli2018}, following the standard steps plus an extra step of GALFIT \citep{Peng2010} modeling of the contamination from more than 80 foreground stars and cluster galaxies.  Since the Galactic coordinates of the Sunburst Arc are close to the Galactic plane, there are more foreground stars in the field of view than is typical for lensed galaxies. Most of the grism spectra are thus contaminated by foreground star lights at several locations. In this paper, we focus on grism spectrum of LyC emitting knot 3 which is devoid of any contaminant, allowing reliable measurement of both continuum and emission lines.
The continuum and emission lines in the spectra were fit as described in \citet{Rigby2021}.

\subsection{Magellan/MagE optical spectroscopy}
Observations of the Sunburst Arc with the MagE spectrograph \citep{Marshall2008} were made as a part of the
extension to \megasauralong\ (\megasaura ).  The main \megasaura\ survey is described
in \citet{Rigby2018}; the \megasaura\ extension will be described in Rigby et al.\ (in prep.) 
Here we summarize key observational details. Observations were conducted  with MagE mounted on the Baade Magellan telescope. Target acquisition was performed by offsetting from a nearby bright star, with acquisition verified from the slit-viewing guide camera. Nine slit positions were obtained; all used the 0.85\arcsec\ slit width except slit position M-0, which used the 1\arcsec\ slit width --- motivated by the atmospheric seeing during the observing runs.  These observations overlap with 9 of the 10 FIRE slit positions. This overlap enables joint analysis of the rest-frame ultraviolet and optical spectra. The MagE slit positions corresponding to FIRE positions are given in Table 1. 

The MagE spectra cover observed wavelengths of 3200 to 8280~\AA. For the Sunburst Arc at z=2.37, this results in rest-frame wavelength coverage of 950~\AA\ to 2457~\AA.  The spectral resolution, as measured from the widths of night skylines, was $5300 \pm 300$ for the observations with the 0.85\arcsec\ width slit, and $4700 \pm 200$  for the observations with the 1\arcsec\ width slit.

The data were reduced as described in \citet{Rigby2018}.

\subsection{X-ray Observations}

The field containing the Sunburst Arc was observed with the {\it Chandra X-ray Observatory} under observation ID 20442 (PI: Bayliss). The purpose of these observations was to test for the presence of an X-ray bright active galactic nucleus (AGN) as a possible source of ionizing radiation in the lensed galaxy. The observations consist of a single 39.53 ks exposure with the aimpoint located near the center of the I3 chip in the ACIS-I array in VFAINT telemetry mode. All {\it Chandra} data reductions and analysis were performed using the {\it Chandra} Interactive Analysis of Observations (\texttt{CIAO v4.13}) with \texttt{CALDB v4.9.6}. We reprocessed the data using the \texttt{CIAO chandra\_repro} routine, and filtered the data for flares using the \texttt{lc\_sigma\_clip} function in the \texttt{lightcurves} Python package that is included in \texttt{CIAO}. The resulting ``clean'' integration time is 38.53 ks. We apply an energy filter of 0.5--7 keV, visually identify bright sources in the data and perform an initial background estimation by masking out those visually identified sources. We then run the \texttt{wavdetect} procedure using the initial background estimate to identify point sources in the data. We then mask the point sources and the bright emission from the galaxy cluster responsible for lensing the Sunburst Arc, and use the resulting ``source-free'' data to perform a final measurement of the background statistics. 

\begin{figure*}
\centering
\hspace*{-50pt}
\includegraphics[scale=0.7]{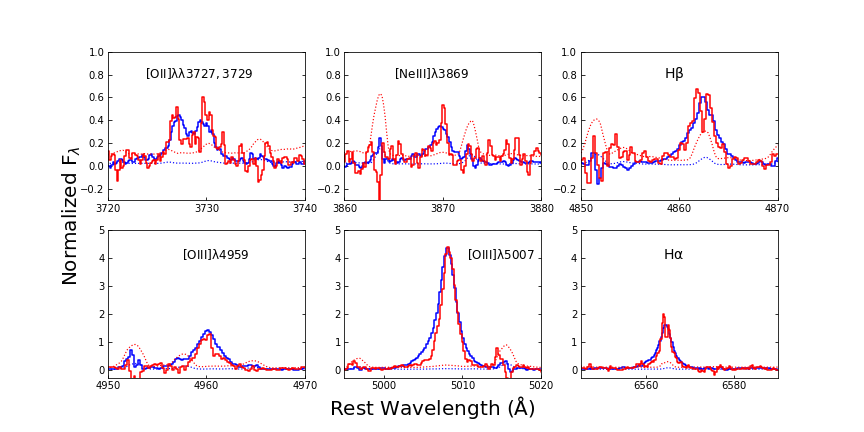}

\caption{Magellan/FIRE spectra (normalized by [OIII]$\lambda$5007) showing key spectral features in the stacked spectrum of LyC leaking regions (blue) and non-leaking regions (red). The corresponding error spectrum is shown in dotted blue (leaker) and dotted red (non-leaker).}
\label{fig:fire_stack}
\vspace*{30pt}
\end{figure*}

\begin{figure*}
\centering
\hspace*{-70pt}
\includegraphics[scale=0.45]{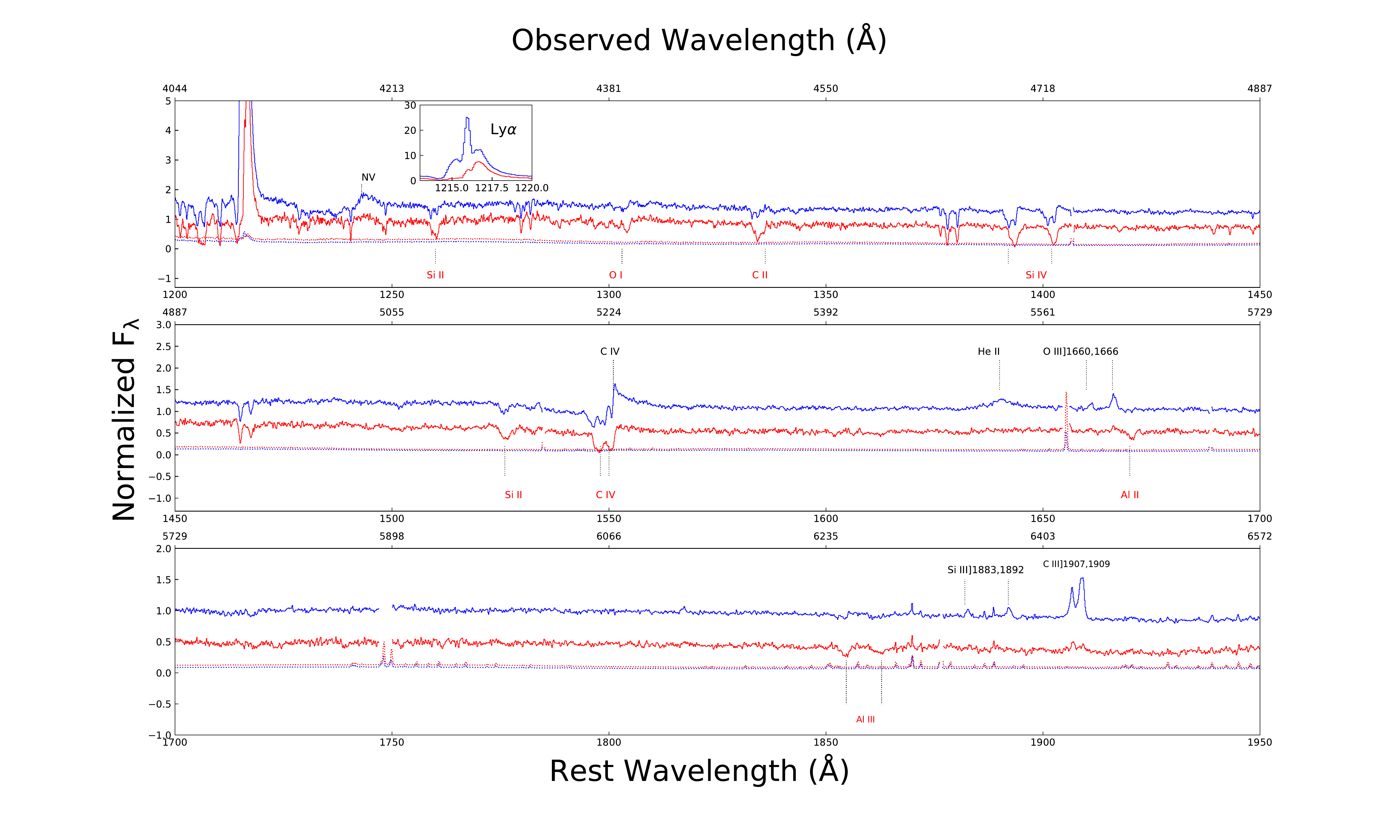}

\caption{Magellan/MagE spectra showing the stacked spectrum of LyC leaking regions (blue) and non-leaking regions (red). The leaker spectrum is normalized by a median continuum level between 1400~\AA\ to 1500~\AA, while the non-leaker is further offset by 0.5 for clarity of plotting. The corresponding error spectrum is shown in dotted blue (leaker) and dotted red (non-leaker). The figure shows several emission (black label) and absorption (red label) features. The emission lines are more prominent in the leaker spectrum while absorption lines, particularly low ionization interstellar features (see  \S3 for more details), are prominent in the non-leaker spectrum.}
\label{fig:mage_stack}
\end{figure*}

\section{Methods}

\begin{table}
\hspace{-2 cm}
\begin{tabular}{lcclcc} \hline
 &  \multicolumn{3}{c}{Leaker} & \multicolumn{2}{c}{Non-Leaker} \\ \hline
  Line & Narrow & Broad & & Narrow & Broad \\ \hline
 $\rm{[OII] 3727}$ & 4.9$\pm$0.2 & 4.5$\pm$0.3 & & 0.6$\pm$0.1 & 0.6$\pm$0.2\\
 $\rm{[OII] 3729}$ &  5.3$\pm$0.2 & 4.8$\pm$0.3 & & 0.8$\pm$0.1 & 0.8$\pm$0.2\\
   $\rm[NeIII] 3869$ & 3.0$\pm$0.2 & 7.5$\pm$0.4 & & 0.5$\pm$0.3 & 0.7$\pm$0.9\\
  H$\beta$  & 6.8$\pm$0.3 & 14.2$\pm$0.4 & & 1.5$\pm$0.2 & 1.5$\pm$0.4 \\
 $\rm{[OIII] 5007}$ & 64.8$\pm$0.3 & 77.8$\pm$0.4 & &  12.2$\pm$0.1 & 4.3$\pm$0.3 \\
 H$\alpha$  & 27.1$\pm$0.5 & 47.1$\pm$0.6 & & 5.6$\pm$0.2 & 3.9$\pm$0.3\\
  $\rm{[NII] 6584}$  & $<$0.6 & $<$1.1 &  & $<$0.3 & $<$0.6\\ \hline
\end{tabular}
\caption{Two component gaussian fits to the stacked FIRE spectra of the LyC leaker and the non-leaker. The table shows line fluxes in both the narrow and broad emission components. Line fluxes are given in the units of 10$^{-17}$ erg s$^{-1}$ cm$^{-2}$.}
\label{table:fire}
\end{table}

\begin{table}
\hspace{-1.2 cm}
\begin{tabular}{lcccc} \hline
MagE slit &  E(B-V) & $\sigma$ & Metallicity (Z$_{\star}$/Z$_{\odot}$) & Age (Myr) \\ \hline
M-0$^{*}$  & $0.146$ & $0.003$ & $0.55$ & $2.92$   \\
M-2$^{*}$  & $0.059$ & $0.003$ & $0.50$ & $3.04$   \\
M-3  & $0.018$ & $0.003$ & $0.23$ & $4.03$   \\
M-4  & $0.078$ & $0.005$ & $0.36$ & $10.1$   \\
M-5  & $0.054$ & $0.004$ & $0.24$ & $12.4$   \\
M-6  & $0.080$ & $0.004$ & $0.28$ & $11.3$   \\
M-7$^{*}$  & $0.098$ & $0.007$ & $0.40$ & $2.60$   \\
M-8$^{*}$  & $0.040$ & $0.002$ & $0.40$ & $2.65$   \\
M-9$^{*}$  & $0.049$ & $0.005$ & $0.44$ & $2.84$   \\ \hline
\end{tabular}
\caption{Derived E(B-V) reddening, stellar metallicity, and stellar age, from stellar population fits
to the MagE data for the Sunburst arc. The asterisk denotes the MagE slits targeting LyC leaking regions.}
\label{table:stellar_pops}
\end{table}

\begin{table}
\begin{tabular}{lcc} \hline
 Line Ratios &  Leaker & Non-Leaker \\ \hline
O32 &  7.5$\pm$0.9 & 6.1$\pm$1.9 \\
$\rm{[OIII]5007}$/H$\beta$ &  6.6$\pm$0.8 & 5.3$\pm$1.5 \\
R23 & 9.4$\pm$1.1 & 8.3$\pm$2.3 \\
$\rm{[NII]6584}$/H$\alpha$ & $<$0.002 & $<$0.009  \\

\end{tabular}
\caption{Table showing optical line diagnostic ratios (combined broad and narrow) for the leaker and non-leaker stacks. The limits are 3-$\sigma$}
\label{table:line_diagnostic}
\end{table}

\subsection{Stacking the MagE and FIRE spectra}
As demonstrated in \citet{Rivera-Thorsen2019}, all the LyC leaking regions in the image plane that were targeted with Magellan/FIRE and Magellan/MagE are in fact multiple images of a single physical region in the source plane.  It is therefore appropriate to combine the spectra from these ``leaker'' slits, since they arise from the same physical source. By contrast, the 4 slits that target non-leaking regions (see Figure~\ref{fig:finder}) correspond, according to the lens model, to at least seven different physical regions. While it is desirable to study these regions separately, for some applications stacking is a practical necessity, as the surface brightness of the non-leaking regions is low. As such, we separately stacked the spectra of LyC leaking regions and the non-leaking regions, to study characteristic differences.

The FIRE stacking was conducted as follows. As Table~1 lists, 6 FIRE slits (F-0, F-1, F-2, F-7, F-8, F-9) cover the LyC leaker, and 4 (F-3, F-4, F-5, F-6) cover non-leaking regions.  Of the 4 slits covering non-leaking regions, we exclude slit F-3 for further analysis since this region may contain an atypical extragalactic object \citep{Vanzella2020}. The remaining spectra are then normalized by their respective [OIII]$\lambda$5007 flux density. 
We then perform average stacking of FIRE spectra of the 6 LyC leaker slit positions and 3 non-leaking slit positions.  Figure~\ref{fig:fire_stack} shows the resulting stacked FIRE spectra.

We also produced stacked rest-frame ultraviolet spectra of the LyC leaker and the non-leaking regions.  All but one FIRE slits (i.e. F-1) have corresponding MagE spectra. We also exclude the MagE spectrum corresponding to the FIRE F-3 position (i.e M-3) for further stacking analysis, resulting in 5 leaker (M-0, M-2, M-7, M-8, M-9) and 3 non-leaker (M-4, M-5, M-6) MagE spectra. Since all LyC leaker images are physically the same star-forming region \citep{Rivera-Thorsen2019}, leaving one leaker region (compared to FIRE stack) without being included in the stacked spectrum other than slightly lowering the signal to noise. We then normalized all the leaker and non-leaker spectra with their continuum flux density from the rest frame 1400~\AA\ to 1500~\AA. After normalization, we perform average stacking to produce the stack rest-frame ultraviolet spectra for leakers and non-leakers. Figure~\ref{fig:mage_stack} shows the stacked rest-frame UV spectra of leaker (blue) and non-leaker (red).

\begin{table*}
\hspace{-2 cm}
\begin{tabular}{lcccccccc}
\hline
 
  Line &   $\rm{[OII]}$  & $\rm[NeIII]+H8+He I$  & [NeIII]+H$\epsilon$   & H$\delta$ & H$\gamma$+[OIII]4363  & H$\beta$  & $\rm{[OIII]4959}$ & $\rm{[OIII]5007}$  \\ \hline
Flux &  19.2$\pm$1.3  & 19.3$\pm$4.0 & 7.5$\pm$2.2  & 7.4$\pm$1.9 & 19.5$\pm$4.8 &  36.6$\pm$2.0  & 80.5$\pm$0.7 & 242.5$\pm$2.1 \\ \hline
EW (\AA)  & 46.7$\pm$3.1  & 52.8$\pm$10.9 & 22.0$\pm$6.5  & 42.1$\pm$6.2  & 74.7$\pm$18.4  & 169.9$\pm$9.3  & 373.4$\pm$3.2 & 1119.7$\pm$9.7 \\ \hline

\end{tabular}
\caption{HST grism measurements of the LyC emitting region (knot 3). Line fluxes and uncertainties are given in the units of 10$^{-17}$ erg s$^{-1}$ cm$^{-2}$.}
\label{table:grism}
\end{table*}
\subsection{Emission line fitting in the FIRE spectra}
The complex emission line profiles shown in Figure 2 require two Gaussian components for each emission line. We fit the observed emission line profiles by assuming that all emission lines can be fit by the same number of components at the same velocity, but with varying total fluxes. We fixed the line flux ratio of [OIII]$\lambda\lambda$5007,4959 to be 3.01 \citep{Storey2000}. We identify [OIII]$\lambda$5007 as the highest signal-to-noise emission line which is not impacted by any sky residual. We fit [OIII]$\lambda$5007 by a two-component Gaussian model, and identify the systemic redshift traced by the narrow component. 

For all other emission lines, we fixed the systemic redshift as well as line widths of two Gaussian components traced by the [OIII]$\lambda$5007 (after accounting for instrument resolution). We also fixed the relative offset between the two components as given by the [OIII]$\lambda$5007 line profile.  However, we set the narrow to broad component integrated flux ratio to be a free parameter. The measured emission line fluxes from the two-component Gaussian model are given in Table 2, for both the leaker and non-leaker stack.

\subsection{Systemic Redshift}
The best published systemic redshift for the Sunburst Arc, $2.3709 \pm 0.001$,  comes from a stellar population fitting to the MagE M-0 spectrum \citep[][]{Chisholm2019}.  The high signal-to-noise of the emission lines in the FIRE data potentially enables a more precise measurement.  We measured the velocity centroids of the narrow components of the brightest line, [OIII]$\lambda$5007, in each FIRE pointing.  The resulting redshifts are given in Table 1. The average redshift from [OIII]$\lambda$5007 emission  is 2.37034$\pm$0.00024, consistent with the previous result. 

\subsection{Stellar Population fitting to the MagE spectra}
Rest-UV spectra contain several key features that indicate the age and metallicity of the massive stellar population. The stellar population fitting of the stacked spectra followed the procedure described in \citet{Chisholm2019}. To summarize: we assume that the spectra can be fit by a linear combination of bursts of stellar populations, each with a given age and metallicity, with a single reddening E(B-V) value. For the theoretical stellar models, we considered single-star Starburst99 models \citep{Leitherer1999, Leitherer2010, Leitherer2014}. We assume a Kroupa initial mass function \citep[IMF;][]{Kroupa2001} with a broken power-law with a high-(low-) mass exponent of 2.3 (1.3) and a high-mass cut-off of 100 M$_\odot$.
We then propagated the intrinsic spectra through a uniform dust screen using the \citet{Reddy2016} attenuation law. We fit the observed spectra between rest frame wavelengths of 1220~\AA\ to 2000~\AA. The outputs of this process  are the linear coefficients for each of the 50 single-age, single-metallicity models, as well as an E(B-V) reddening value.  

\subsection{Lensing Magnification}

We define pixel-level masks that identify the FIRE slit positions along the giant arc from the available {\it HST} imaging by flagging which pixels fall inside or outside of each observed FIRE slit position. Only pixels that have more than half of their area falling within the slit aperture are counted as inside the slit, and any effects from fractional slits should be extremely minor considering the small pixel size ($0.03$\arcsec) in our drizzled {\it HST} images relative to the FIRE slit size. The slit definition masks are then used to define an aperture mask for each position that represents the intersection of the FIRE slit aperture and the emission from the giant arc. These apertures represent {\it HST} pixel-level masks indicating the region on the sky from which light entered the FIRE slit at each position listed in Table~\ref{table:observation}.

For the purpose of defining these apertures we use the {\it HST} F125W (the band closest to the J-band used in the FIRE acquisition camera). We compute the sky background statistics in F125W and manually mask out any other sources that fell into some slits. We then define the emission from the giant arc as originating from all pixels within a given FIRE slit that are 8-$\sigma_{sky}$ above the sky background level. We experimented with several different integer multiples of $\sigma_{sky}$ as the threshold, and found that 8-$\sigma_{sky}$ strikes a good balance between robustly identifying the  giant arc emission without including spurious sky pixels, while also capturing the large majority ($\gtrsim 90$\%) of the total giant arc flux in each slit.

The average magnification within each aperture is determined as follows. We first ray-trace each aperture from the image plane to the source plane using the lens model deflection map outputs. We then calculate the area enclosed by the aperture in the source plane and in the image plane. The average magnification is defined as the ratio of the image plane to source plane areas. 

Some of the apertures cross the critical curves and include portions of the arc that traverse it, causing the source-plane projection of the aperture to fold over itself \citep[see, e.g., Figure~6 of][for an illustration of this behavior]{sharon12}. 
Therefore, care must be taken in deriving their average magnification. In such cases, the overlapping region in the source plane should only be counted once and the average magnification accounts for the fact that the slit aperture encloses two images of the same source region. 

Magnification uncertainties are estimated by running the same process on deflection maps calculated from sets of parameters in the MCMC that sample $1-\sigma$ in the parameter space. The resulting magnifications and their uncertainties are tabulated in Table~\ref{table:observation}.

\section{Results}

\subsection{Nebular Reddening} 
We applied a nebular reddening correction using the Balmer lines (H$\alpha$ \& H$\beta$) measured in the leaker stack spectrum. While H$\beta$ is slightly impacted by an underlying skyline in the individual spectra, we  successfully recover it in the stack spectrum. We used the observed H$\alpha$/H$\beta$ line ratio of 3.533, and first applied Milky Way extinction correction using E(B-V)=0.094 \citep[][]{Green2015}. In the next step, we assumed Case B recombination to infer intrinsic value of $(H\alpha/H\beta)_{in}$=2.86 corresponding to a temperature of $T=10^4 $K and an electron density of $n_{e}=10^2$ cm$^{-3}$. For an extinction law of \citet{Cardelli1989} and an R$_V$ value of 3.1, this corresponds to a nebular reddening of E(B-V)=0.195$\pm$0.025. Recovering reliable H$\beta$ measurement in the non-leaker stack is challenging for two reasons. First, non-leaking regions have lower surface brightness and hence lower single-to-noise in individual emission lines. Second, only 3 spectra are stacked for non-leaker (compared to 6 in leaker), again resulting in lower single-to-noise emission lines. This prevents us from estimating a reliable H$\alpha$/H$\beta$ ratio for the non-leaker. We, therefore, assume that the nebular reddening value of the leaker is representative of the whole Sunburst Arc. 

\subsection{Stellar Reddening} 
The stellar reddening is inferred by fitting the rest-frame UV MagE data (see \S2.4).  These results are given in Table~\ref{table:stellar_pops}, and cover a larger range of physically distinct regions than is currently available for the reddening data.  We, therefore, use this stellar reddening information to check whether there are large spatial gradients in the reddening.  
The results presented in Table~\ref{table:stellar_pops} show no apparent trend of $E(B-V)_*$ with whether the slit covers the leaking or non-leaking regions.  This supports the validity of using a single nebular reddening value for the whole of the Sunburst Arc.

\begin{figure*}
\centering

\hspace*{-20pt}
\includegraphics[scale=0.68]{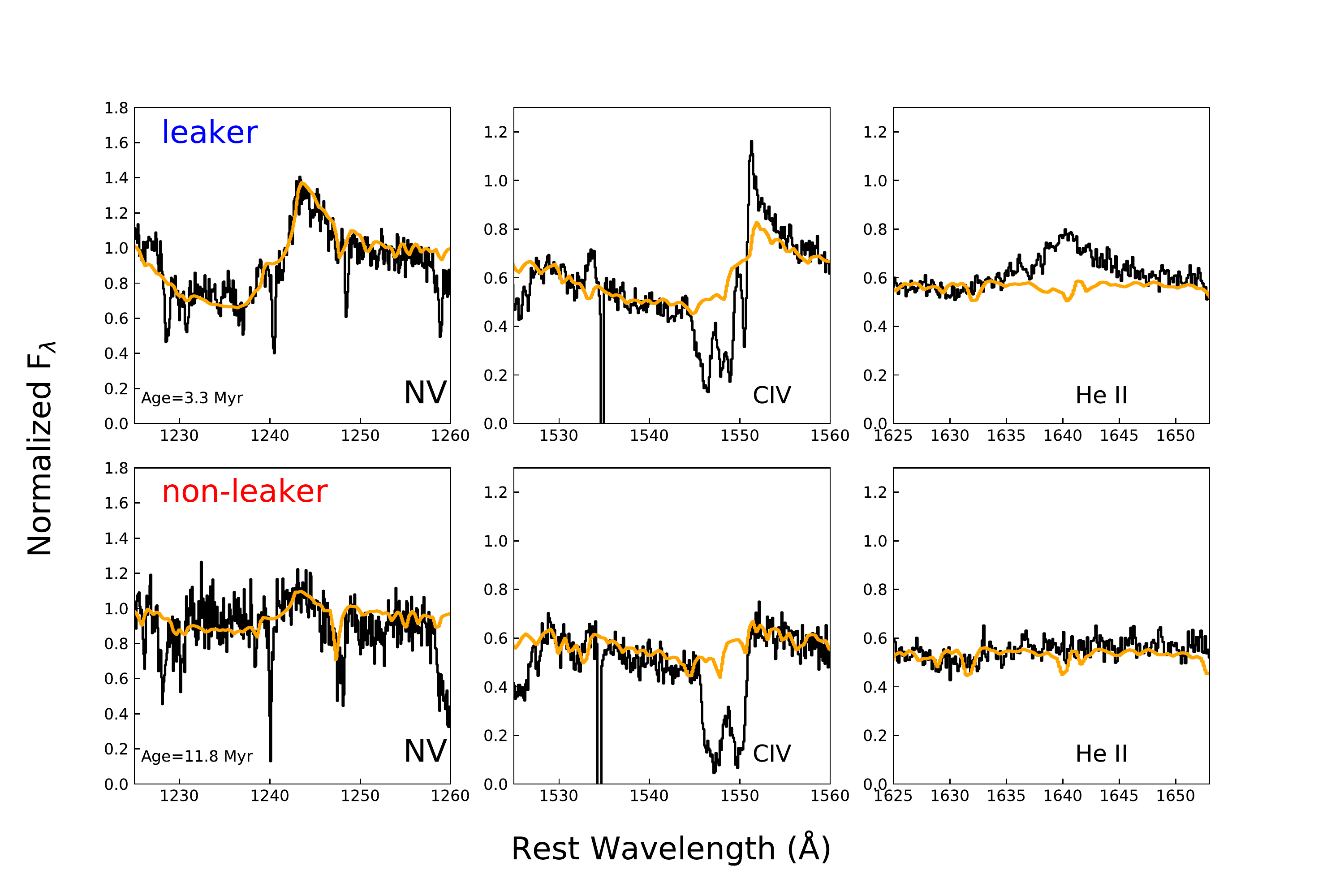}
\caption{Stellar population synthesis fit to the leaker and non-leaker spectra. The left panel shows P-Cygni NV features in the leaker (top) and non-leaker (bottom), the middle panel shows P-Cygni CIV features in the leaker (top) and non-leaker (bottom) and the right panel shows He II 1640 feature in the leaker (top) and non-leaker (bottom). The orange line represents the best-fit light-weighted SB99 model: the inferred age is 3.3 Myr for the leaker and 11.8 Myr for the non-leaker. }
\label{fig:stellar_age}
\end{figure*}

\begin{figure}
\centering

\hspace*{-30pt}
\includegraphics[scale=0.38
]{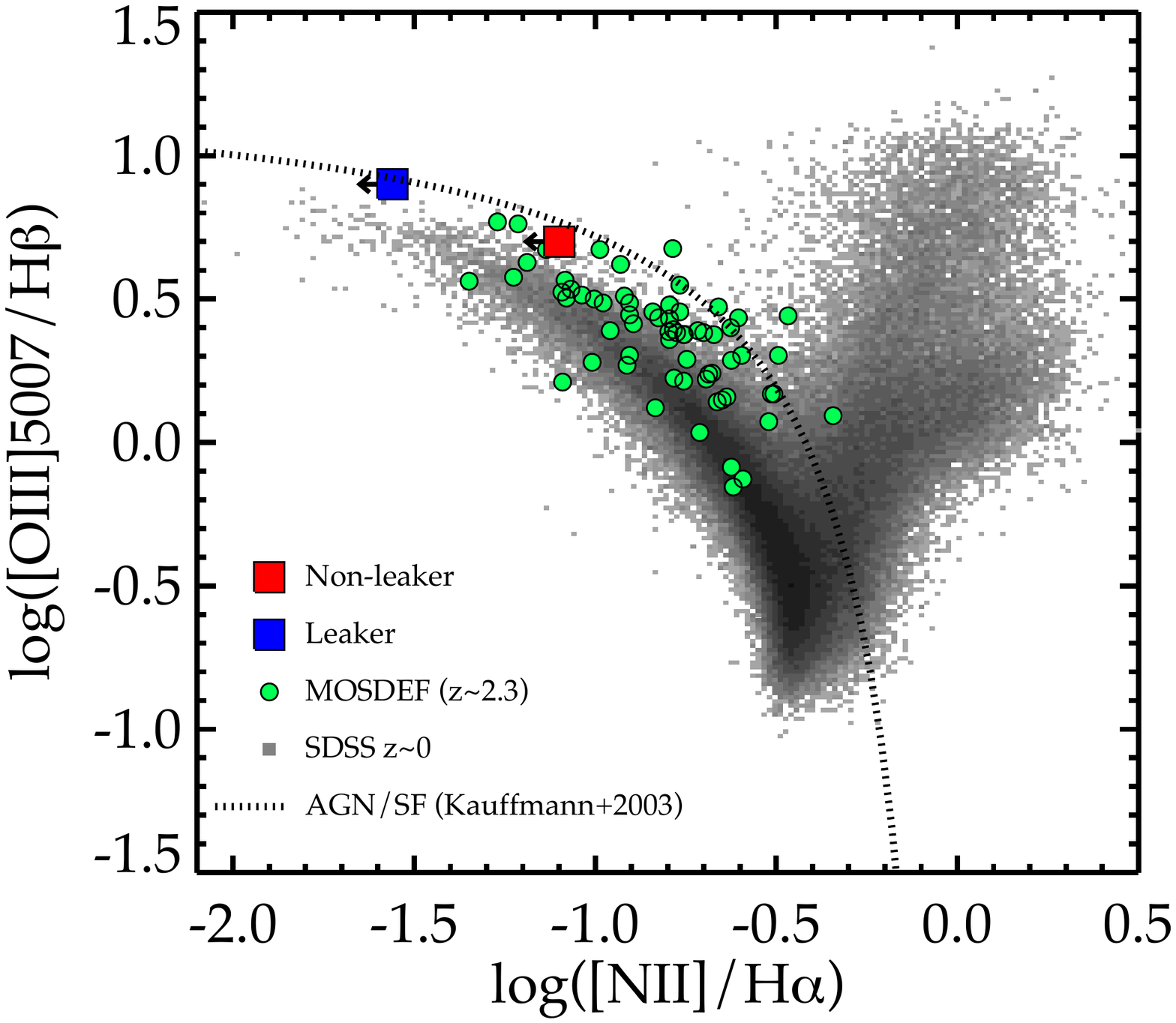}

\caption{The ionization diagram showing the location of leaker (blue) and non-leaker (red) using combined broad and narrow components. The grey squares represent SDSS data points and green circle are z$\sim$2.3 galaxies from MOSDEF survey \citep{Sanders2016}. The dotted line represents AGN/SF demarcation from \citet{Kauffmann2003}.}

\label{fig:bpt}
\end{figure}

\subsection{Comparison of UV features in leaking and non-leaking regions }
\label{uv_features}
Figure~\ref{fig:mage_stack} plots the MagE spectra for the leaker and the non-leaker.  The two spectra are quite different in several ways.  First, as shown in the figure inset, the Ly$\alpha$ velocity profiles are radically different. The leaker shows a triple-peaked profile which is a characteristic feature of ionizing photons leaking from the galaxy. This feature is clearly absent in the non-leaker stack spectrum. More specifically, the non-leaker lacks the blue-shifted Ly$\alpha$ emission, which would suggest that the line of sight ISM is sufficiently neutral to absorb Ly$\alpha$ photons.  This confirms, with deeper data, the results seen by \citet{Rivera-Thorsen2017}.

Second, the rest-frame UV nebular emission lines, namely CIII] $\lambda\lambda$ 1907, 1909, OIII] $\lambda\lambda$1661, 1666, Si III] $\lambda\lambda$1883, 1892, have much larger equivalent widths in the leaker stack than the non-leaker stack. These features are characteristics of galaxies dominated by young massive star populations \citep{Stark2013b, Berg2016, Du2018, Rigby2021}. 

Third, the spectral wind features that are characteristics of massive stars are much stronger in the leaker than in the non-leaker.  These features are described in detail in \citet{Chisholm2019}.  
Figure~\ref{fig:stellar_age} compares these features for the leaker and the non-leaker, namely NV, C~IV, and He II emissions. In particular, the leaker shows strong P-cygni NV and C~IV emission, which is characteristic of an extremely young (2--5 Myr) stellar population. In contrast, these features are very weak in the non-leaker (Figure~\ref{fig:stellar_age}). The leaker shows CIV emission, both a narrow component and a broad component which again is characteristic of massive star winds. The narrow component is likely nebular emission, which is commonly detected in LyC emitting sources \citep[e.g.][]{Naidu2022, Schaerer2022}. CIV also shows the blue-shifted absorption that is characteristic of stellar winds.  These features are much weaker in the non-leaker.  
Additionally, the leaker shows broad He II emission that is entirely absent in the non-leaker. This is only found in massive Wolf-Rayet stars and must indicate that there is a very recent burst of star formation.
Together, these spectral features indicate a very different light-weighted age for the leaker stack compared to the non-leaker stack, which we now quantify.

\begin{figure*}
\centering

\hspace*{-70pt}
\includegraphics[scale=0.45]{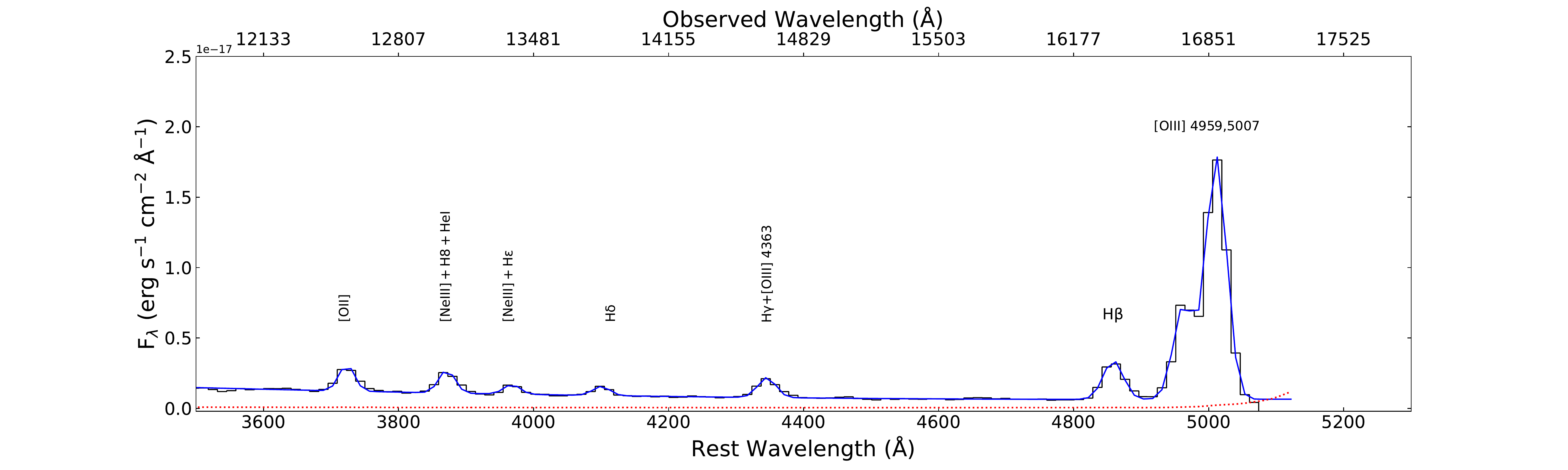}
\caption{HST WFC3/IR G141 grism spectrum of a LyC leaking region (knot 3) in Sunburst Arc. The black line represents grism spectrum, the red dotted line shows noise level and the blue line shows the best fit to the data.}
\vspace*{30pt}
\label{fig:grism}
\end{figure*}

\begin{figure*}
\centering

\hspace*{-20pt}
\includegraphics[scale=0.78]{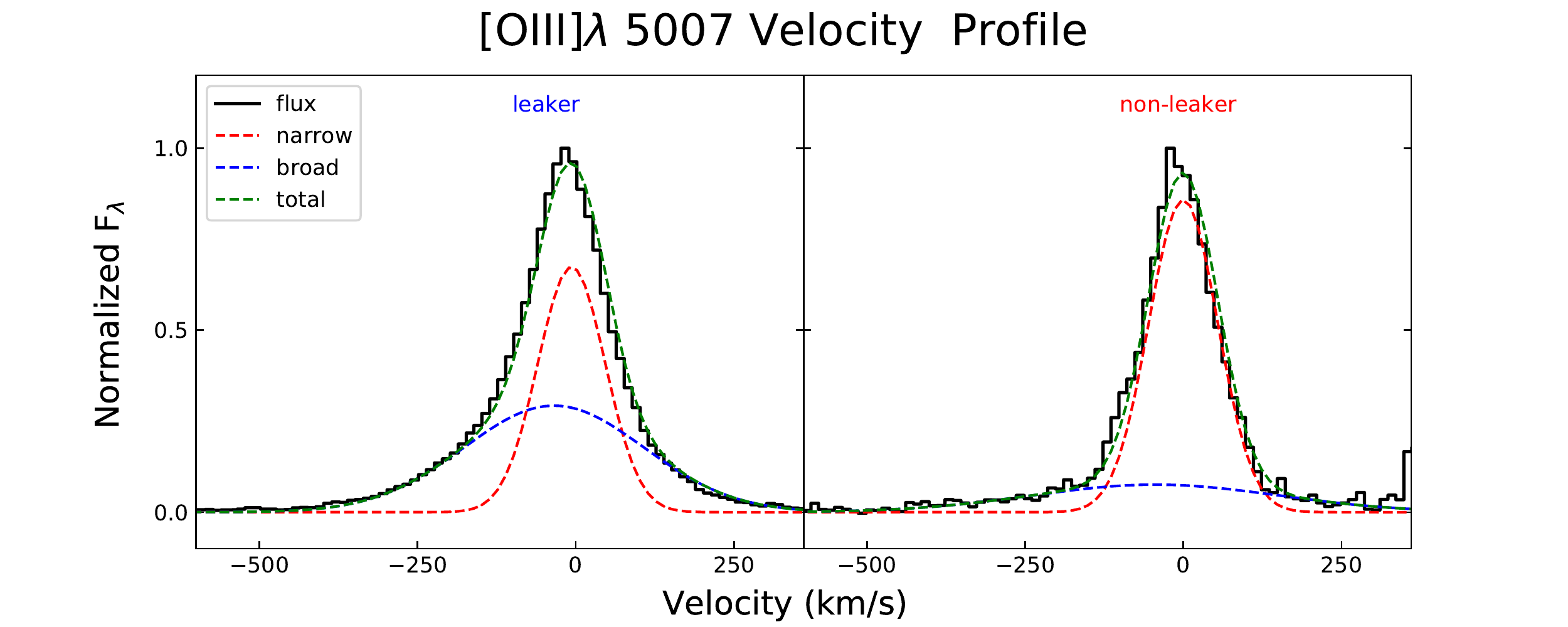}

\caption{[OIII]$\lambda$5007 emission line profile in leaking (left) and non-leaking (right) regions. The black curve represents FIRE spectra. We fit the data with a two-component gaussian model where the red dashed line represents the narrow component and the blue dashed line represents the broad line. The green dashed line represents best-fit model.}
\label{fig:oiii_profile_compare}
\end{figure*}

\subsection{Stellar age and metallicities}
The stellar population synthesis fits use both the strong stellar wind lines (such as N~V and C~IV) and the weak stellar photospheric absorption features to match the observations as a best-fit stellar templates. Using this information, the fitting estimates a UV light-weighted age of  $3.3 \pm 0.5$ Myr and a metallicity  of 0.46$\pm$0.06 Z$_{\odot}$ for the leaking region, and $11.8 \pm 0.9$ Myr and 0.30$\pm$0.05 Z$_{\odot}$ for the non-leaking region. 

As discussed in Section~\ref{uv_features}, Figure~\ref{fig:stellar_age} compares the observations (in black) to the data (in gold) for the leaker (upper panel) and non-leaking (lower panel) spectra. The left and middle panels show the stellar population fits to the strong N~V and C~IV stellar wind lines. These wind lines display the classic P-Cygni wind feature with blue shifted absorption and redshift emission arising from gas launched off the photospheres of massive stars. While both the leaking and non-leaking fits match the absorption components, the C~IV feature for the leaker shows evidence of narrow C~IV emission that is not well-fit by the model. The He~II region is well-fit in the non-leaker case, but poorly fit in the leaker case. This is due to broad ($\sim2000$~km~s$^{-1}$) emission seen in the leaker spectrum. Broad He~II emission is typically seen in massive, moderately metal-rich Wolf-Rayet stars with very young ages. Thus, the observed broad He~II is consistent with the young estimated age of the leaker region, and the absence of broad He~II in the non-leaking region is consistent with the older fitted stellar population. The fact that the Starburst99 models do not match the He~II feature is a common problem with the current generation of stellar models because the stellar tracks do not populate the Wolf-Rayet stars that produce this feature \citep{Leitherer2010, Chisholm2019}. Thus, the dichotomy of stellar properties -- young and moderate metal-rich population in the leaking region and a older population in the non-leaking region -- is consistent with the observed stellar features.

\subsection{Rest-frame optical diagnostic line ratios}
The rest-frame optical strong emission lines form the classic line ratio diagnostics that are used to classify galaxies as dominated by star formation or active nuclei, and are sensitive to the metallicity and ionization state of the nebular gas. In Table~\ref{table:line_diagnostic} we report the measured values for the standard emission line ratios (for combined broad and narrow components) after accounting for the internal extinction discussed above. Both the leaker and non-leaker show larger than average O32 and [OIII]5007/H$\beta$  of typical $z\sim2$ galaxies suggesting that the galaxy is in higher ionization state compared to typical galaxies at similar redshift \citep{Steidel2016,Sanders2016}. O32 is often thought of as a proxy of LyC escape \citep[e.g.][]{Jaskot2013,Nakajima2014}, but is often not seen to scale with LyC escape at low-redshift \citep{Izotov2008, Flury2022a}. Interestingly, these line ratios are not very different for individual leaker and non-leaker stacks. We note that the non-leaker stack doesn't include all non-leaking regions identified in the Sunburst Arc.   

In Figure~\ref{fig:bpt}, we plot the classic BPT diagram \citep{Baldwin1981} in [O~III]/H$\beta$ vs [N~II]/H$\alpha$ plot. The grey data points indicate SDSS galaxies (z$\sim$0) and the green circles indicate z$\sim$2.3 galaxies from the MOSDEF survey \citep[e.g.][]{Sanders2016}, while the dotted line indicates AGN versus star-forming galaxies classification based on \citet{Kauffmann2003} model. The line flux ratios of both leaker (blue) and non-leaker (red) are shown in the diagram. Since [N II] is not detected in either stack, we only have limits on the [N~II]/H$\alpha$ ratio, but these limits tightly constrain both the leaking and non-leaking portions of the Sunburst Arc to the far upper left wing of the ``seagull''--shaped cloud that SDSS galaxies inhabit.  Both leaker and non-leaker stacks show high [OIII]/H$\beta$, distinct from typical star forming galaxies in SDSS sample. However, such a high [OIII]/H$\beta$ is consistent with young metal-poor galaxy population typically observed at the intermediate redshifts (e.g. \citealt{Berg2018, Vanzella2017, Mainali2020}).

\subsection{X-ray constraints on AGN}

We search for X-ray emission from the lensed Sunburst Arc to look for direct evidence of any possible active galactic nucleus (AGN) that could be contributing substantially to the ionizing radiation in the galaxy. We define an aperture for the {\it Chandra} data by convolving the {\it Chandra} PSF with contours that trace the optical emission of the Sunburst Arc in {\it HST} imaging. The bright, extended X-ray emission from the foreground lensing galaxy cluster overlaps with the location of the lensed images of the Sunburst Arc, effectively acting as another source of background noise for the purpose of constraining the emission from the arc. To account for the X-ray photons from the cluster we fit a radially symmetric $\beta$-profile to the cluster emission. We use the model to subtract off the cluster emission underneath the giant arc aperture, resulting in a measurement of the X-ray emission from the Sunburst Arc; the result is a non-detection of the giant arc, with a 0.5--7 keV 2-$\sigma$ limiting flux of $6.6 \times 10^{-15}$ erg s$^{-1}$ cm$^{-2}$. Assuming a power-law spectra shape with $\Gamma = 1.8$ (reasonable for AGN), this corresponds to a rest-frame 2--10 keV flux limit of $3.6 \times 10^{-15}$ erg s$^{-1}$ cm$^{-2}$ at $z=2.37$, and an upper limit on the intrinsic 2--10 keV luminosity of L$_{\rm 2-10} < 1.6 \times 10^{44}$ erg s$^{-1}$. These limits do not account for the lensing magnificiation, which reduces the constraint on the upper limit by the magnification, $\mu$. We estimate the median magnification factor based on our strong lensing model of the Sunburst Arc system using the full aperture within which the {\it Chandra} upper limit was measured; the resulting magnification is $\mu \simeq 38$. Applying this magnification factor to the upper limit on the X-ray luminosity constrains the intrinsic X-ray luminosity of the Sunburst Arc to be L$_{\rm 2--10} \lesssim 7.6 \times 10^{42}$ erg s$^{-1}$, ruling out the presence of an X-ray bright AGN in the Sunburst Arc.

\begin{table*}
\hspace{-1 cm}
\begin{tabular}{lccccc}
\hline
 
  & Centroid (Narrow) & FWHM (Narrow) & Centroid (Broad) & FWHM (Broad) & Fractional light in broad component \\
    & km s$^{-1}$ & km s$^{-1}$ & km s$^{-1}$ & km s$^{-1}$ & \% \\
Leaker & 0 & 112$\pm$1 & -27$\pm$1 & 326$\pm$2 & 54.5$\pm$0.3 \\ \hline
Non-leaker & 0 & 115$\pm$2 & -43$\pm$14 & 454$\pm$47 & 26.1$\pm$7.0 \\ \hline

\end{tabular}
\caption{Breakdown of narrow and broad components in leaker vs non-leaker measured from [OIII]$\lambda$5007 emission.}
\label{table:leaker_vs_nonleaker}

\end{table*}

\begin{table}
\begin{tabular}{lcc}
\hline
 
 FIRE slits & Physical regions & Distance (kpc) \\

  \hline
F-0  & 1 & 0  \\ \hline
F-4  & 3,4,7,9,10 & 1.53$^{+0.50}_{-0.23}$ \\ \hline
F-5  & 3,8 & 1.06$^{+0.20}_{-0.11}$  \\ \hline
F-6  & 2 & 0.47$^{+0.03}_{-0.07}$ \\ \hline

\end{tabular}
\caption{Distinct physical regions of Sunburst Arc targeted by FIRE spectra \citep{Sharon2022}. From left to right, the columns are: FIRE slit name, distinct physical regions within the FIRE pointing and average distance (pc) of the physical regions from LyC emitting region in source plane}
\label{table:physical_regions}

\end{table}

\begin{table}
\begin{tabular}{lcc}
\hline
 
Diagnostic & leaker (cm$^{-3}$)  & non-leaker (cm$^{-3}$) \\

  \hline
CIII]$\lambda\lambda$1907,1909 & 66000$^{+5000}_{-3000}$ & 19000$^{+7000}_{-4000}$   \\ \hline
[OII]$\lambda\lambda$3727, 3729 & 335$^{+20}_{-21}$ & 340$^{+1040}_{-340}$ \\ \hline

\end{tabular}
\caption{Electron density measurements for the leaker and the non-leaker using [OII] and CIII] doublets.}
\label{table:leaker_vs_nonleaker_ne}

\end{table}


\subsection{Equivalent widths of the rest-frame optical emission lines}
The equivalent width of nebular emission lines (EW) tells us about the relative strength of the nebular emission compared to the stellar emission. Because it is a ratio, it is not affected by lensing magnification.  It therefore tells us about the intensity and age of the massive stellar population. The non-detection of continuum in the FIRE data precludes us from a reliable equivalent width measurement, but we can measure them using the grism data.

We plot the leaker spectrum in Figure~\ref{fig:grism} and present the measurements in Table~\ref{table:grism}, whereas a similar measurement for non-leaker is currently not possible with the grism data. This is because continuum is weakly detected in the non-leaker owing to its lower surface brightness and is further contaminated by nearby starlights. For the leaker, the equivalent widths are very high, which is typical of LyC emitting galaxies \citep[e.g.][]{Izotov2018, Flury2022b}. Such large equivalent widths are also commonly observed among extreme emission line galaxy populations \citep{Mainali2020, Amorin2017, Labbe2013, Berg2021}.

\subsection{Evidence for ionized outflows in the stacked spectra}
The strongest optical emission line in the FIRE data is [OIII] $\lambda$5007, which is also free of contamination from telluric skylines.  Examining Figure~\ref{fig:fire_stack}, it is immediately obvious that the stacked leaker and stacked non-leaker FIRE spectra show very different velocity profiles for [O III]~5007.  The leaker stack shows a strong broad wing of emission, which is considerably weaker in the non-leaker stack. The leaker stack also shows a broad red wing.  

Table~\ref{table:fire} reports the relative fluxes of the broad and narrow components for both the leaker stack and the non-leaker stack.    
Figure~\ref{fig:oiii_profile_compare} shows in detail the best-fit broad and narrow components of the [O III] $\lambda$5007 velocity profile.  
The observed line widths (FWHM) of the narrow component and broad components in the leaker are $\sim$112 km s$^{-1}$ and $\sim$326 km s$^{-1}$. The broad emission is blue-shifted from the narrow component by $\Delta V$ = 27 km s$^{-1}$. 
For [O III] 5007, 54.5$\pm$0.3\% of the line flux is in the broad component for the leaker stack; and 26.1$\pm$7.0\% for the non-leaker stack. The measurements are given in Table \ref{table:leaker_vs_nonleaker}.

Though it suffers from some skyline contamination (particularly in the non-leaker stack), the H$\alpha$ velocity profiles tell the same story: the leaker stack shows a broad blue wing that is much stronger than in the non-leaker stack.  H$\alpha$ does appear to have a higher fraction of the flux in the broad component compared to [OIII] 5007 line (63.4$\pm$1.5\% \& 41.1$\pm$9.3\% for leaker and non-leaker, respectively).

\subsection{Evidence for ionized outflows in the individual spectra}
The above analysis was performed on the stacked FIRE spectra.   Each spectrum that makes up the stacked leaker spectrum is intrinsically the same source, multiply imaged.  There is therefore no value in analyzing the kinematics of each individual leaker spectrum.  By contrast, the non-leaker stack is comprised of six different physical regions based on the lens model \citep{Sharon2022}, even though only four FIRE slits target non-leaking regions. This is because a FIRE slit may target several unresolved physical regions. The leaker FIRE spectrum F-0 targets Clump 1 while non-leaker spectrum F-4 targets Clumps 2,3,4,9, and 10, F-5 targets Clumps 3,8, and 10 and F-6 targets Clump 2 (Table \ref{table:physical_regions}). We therefore further examine the [O III] 5007 in the individual (non-stacked) non-leaker FIRE spectra to study any spatial variation in the line profile. We calculated angular separation in object planes for different star-forming clumps in the leaker and non-leaker spectra. Then we converted them to average distance (in parsec) from the leaking region to regions targets by F-4, F-5 and F-6 (see Table 6). 

\begin{figure}
\centering

\hspace*{-10pt}
\includegraphics[scale=0.38]{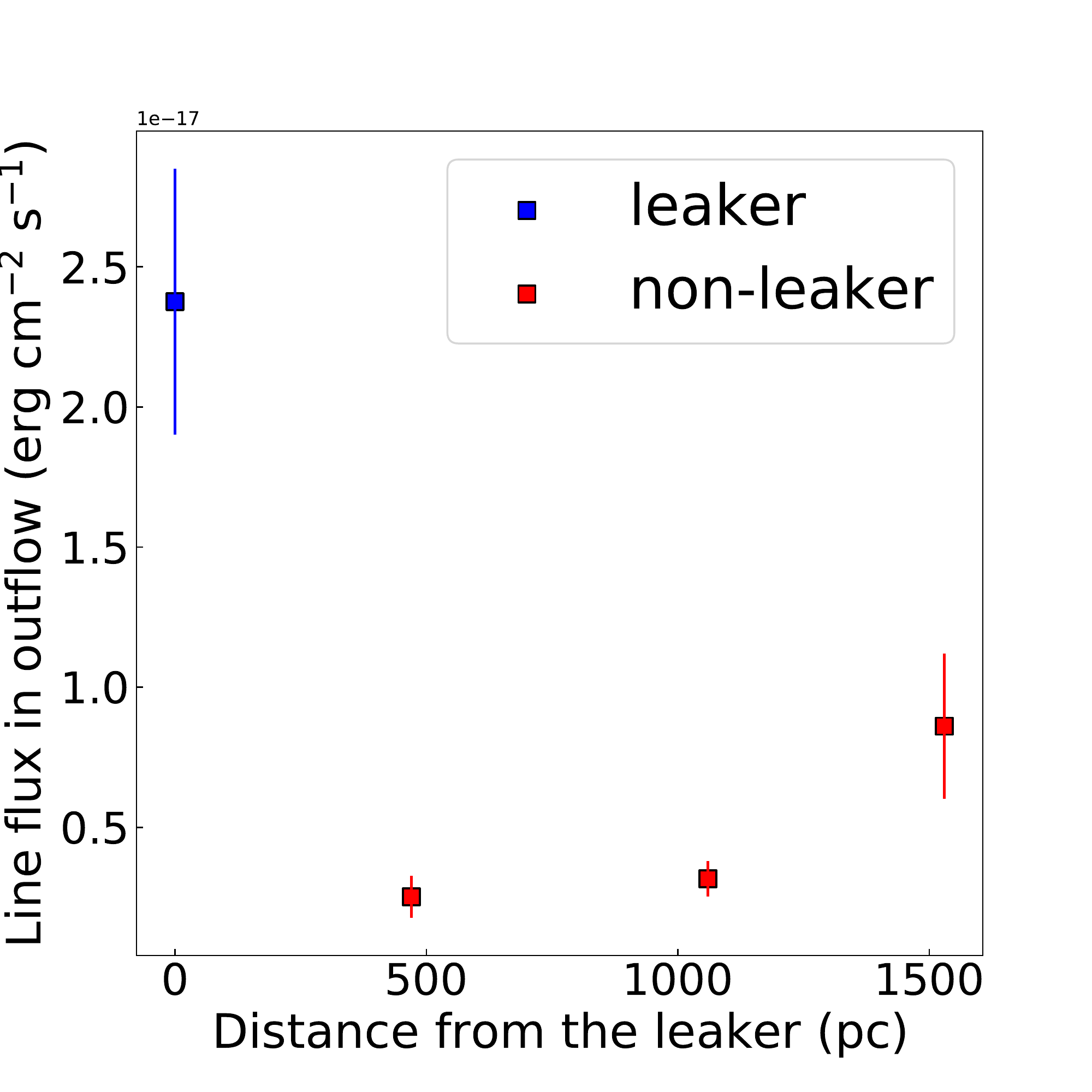}
\caption{Intrinsic line flux (magnification corrected) of outflow component measured from the broad component in [OIII] 5007 line for the leaker (square) and three individual non-leakers (red square) probed by FIRE slits. In the x-axis, we plot physical distances between the leaker and  the three non-leakers in the source plane as given by the lens model \citep{Sharon2022}. }
\label{fig:intrinsic_outflow}
\end{figure}

In Figure~\ref{fig:intrinsic_outflow} we show the intrinsic line flux in the broad component of [OIII] 5007 line as a function of the average distance from the leaker. The intrinsic line flux is obtained after applying magnification correction presented in Table~\ref{table:observation} to the observed line fluxes.  Clump 2 (at 400 pc from the leaker) lies very close to the critical curve in the source plane resulting in extremely large magnification ($\mu$=637). This implies intrinsically negligible flux contribution from Clump 2. While we see a broad outflow component in every physical region, the figure demonstrates that the ionized gas outflow is strong in the leaker, and that it is weak at the 0.5 -- 1 kpc distances probed by the FIRE pointings that cover non-leaking regions.

\subsection{The outflow as traced in absorption of low-ionization and high-ionization gas}
The rest-UV stacks shown in Figure~\ref{fig:mage_stack} show several interstellar absorption features which tell us about the line of sight absorbing gas kinematics. These absorption features are blue-shifted with respect to the systemic velocity which indicates the presence of outflowing gas, consistent with outflows traced in the emission lines discussed above. While flux in the emission lines depends on n$_{e}^2$, the outflows traced by line of sight absorption scales as n$_{e}$. 
Absorption is more sensitive to gas at larger distances, and gas with lower densities. Thus, the absorption lines probe different physical regions of outflow than the emission lines. The absorption lines also have sensitivity to a range of ionization states. 

\begin{figure*}
\centering

\includegraphics[scale=0.5]{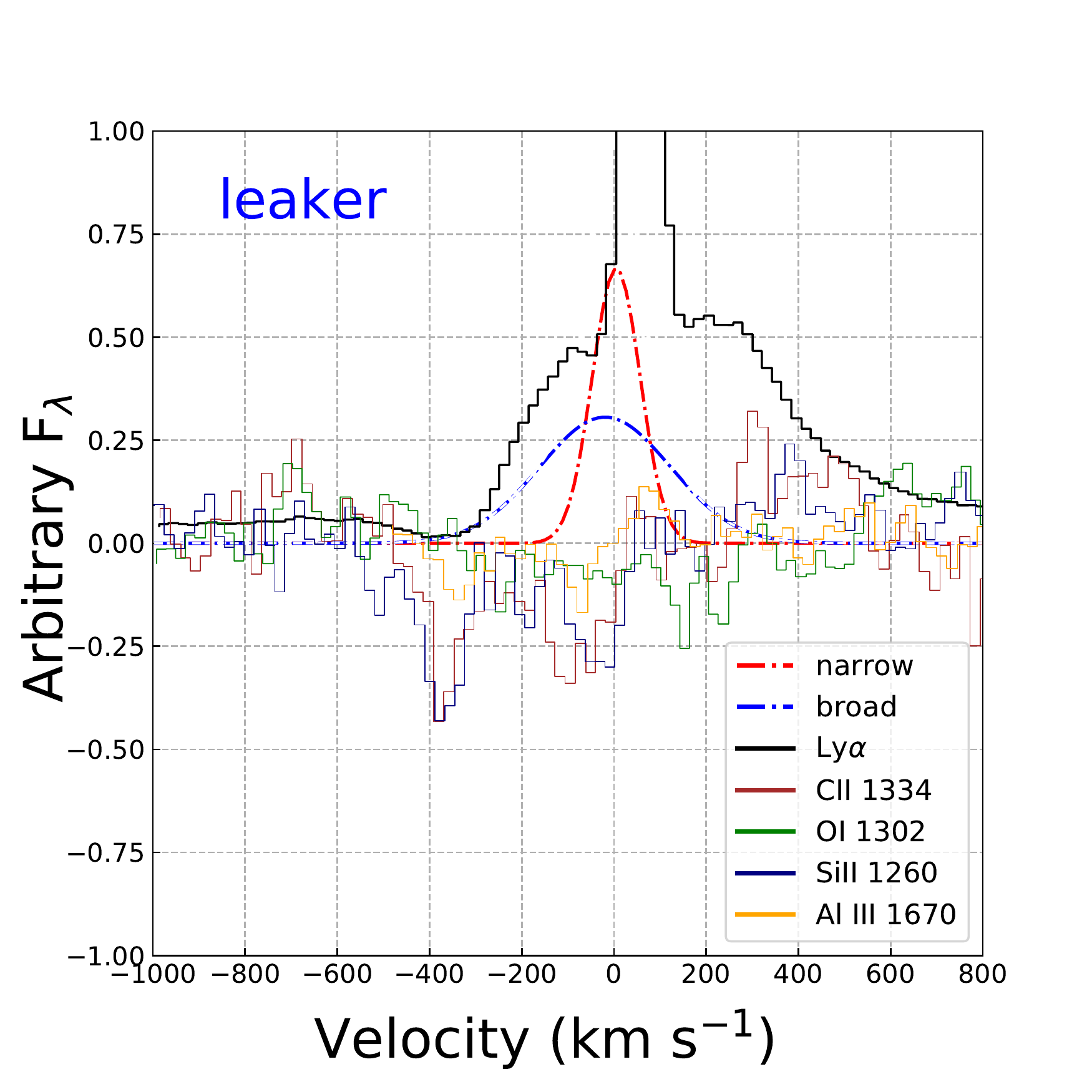}
\includegraphics[scale=0.5]{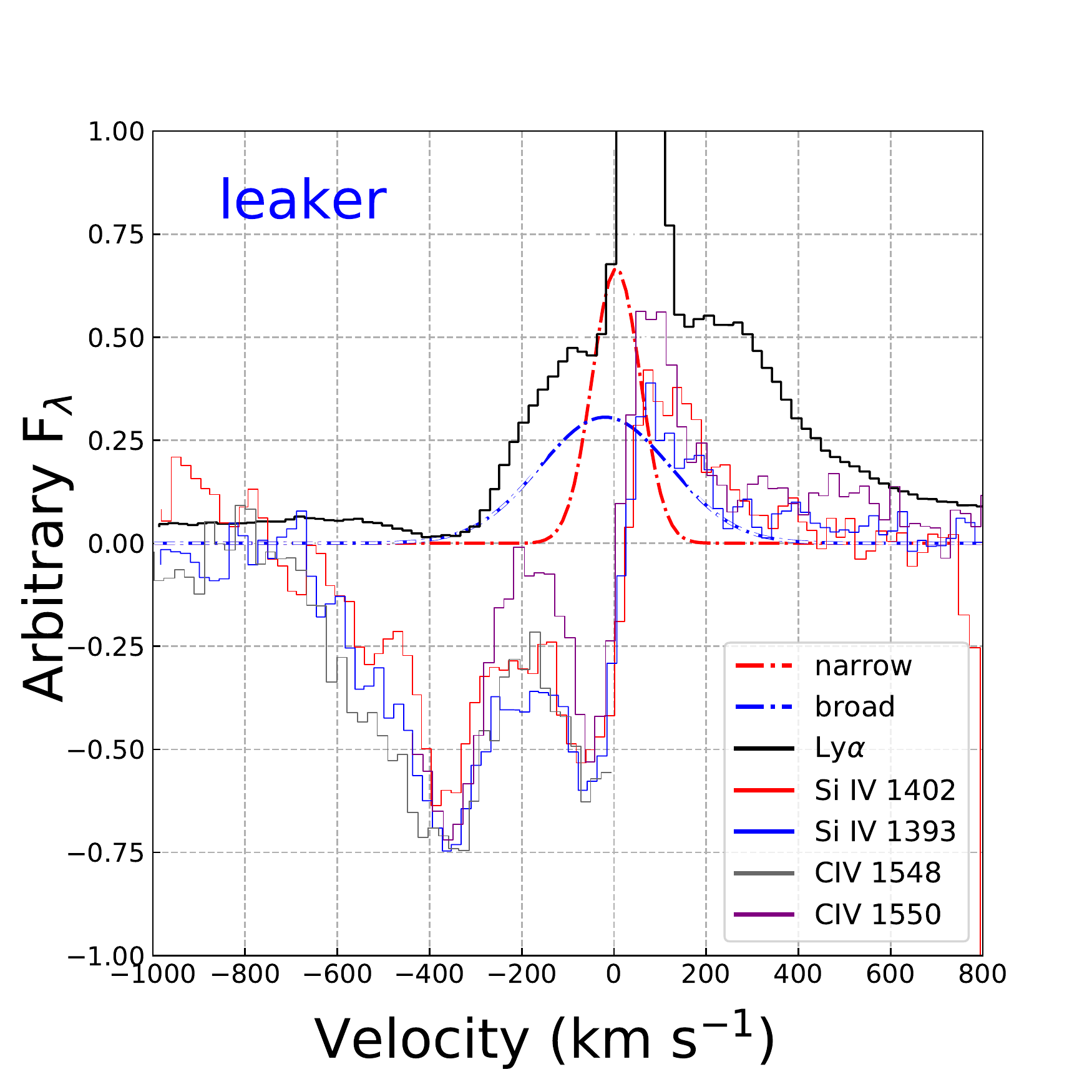}\\
\vspace{-25pt}
\includegraphics[scale=0.5]{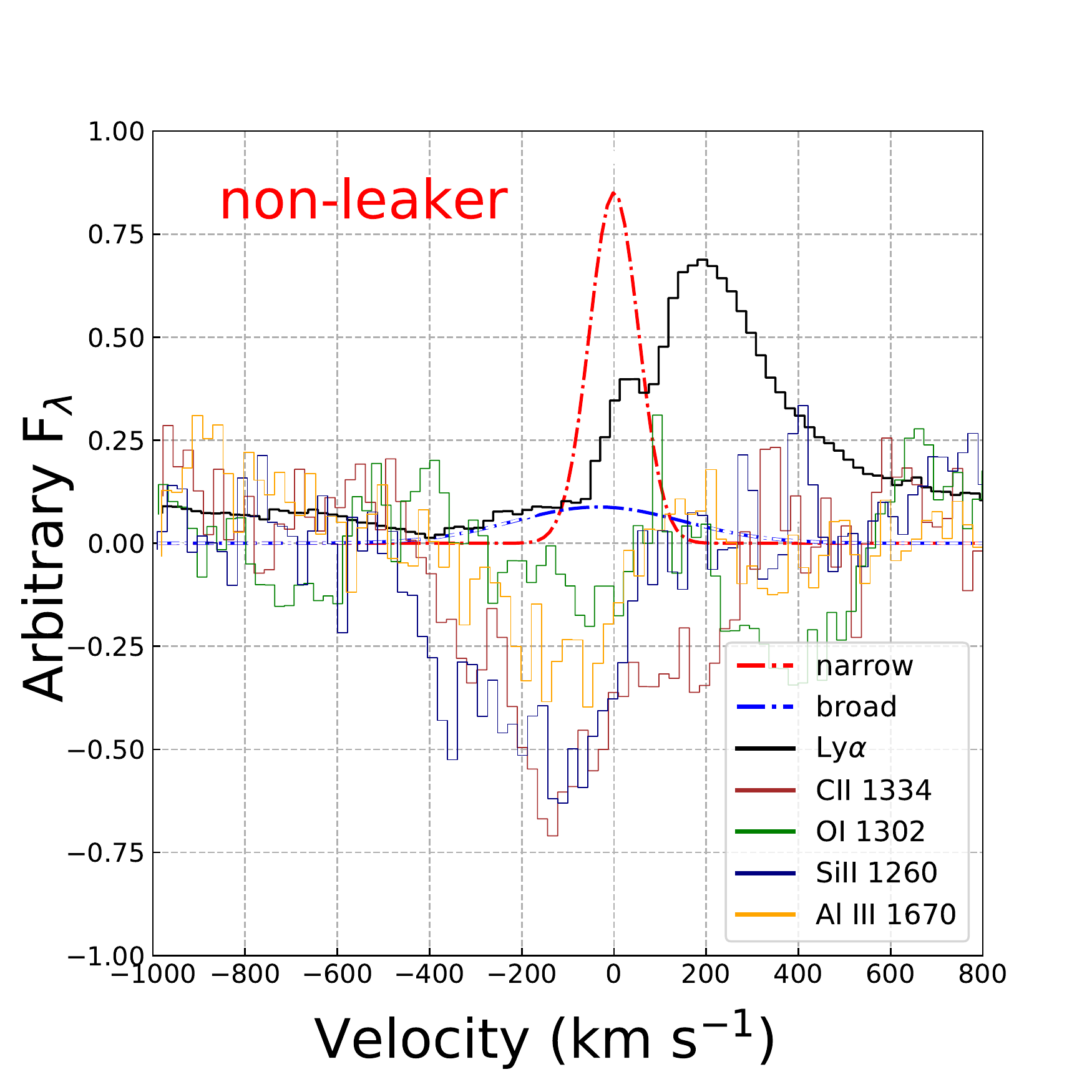}
\includegraphics[scale=0.5]{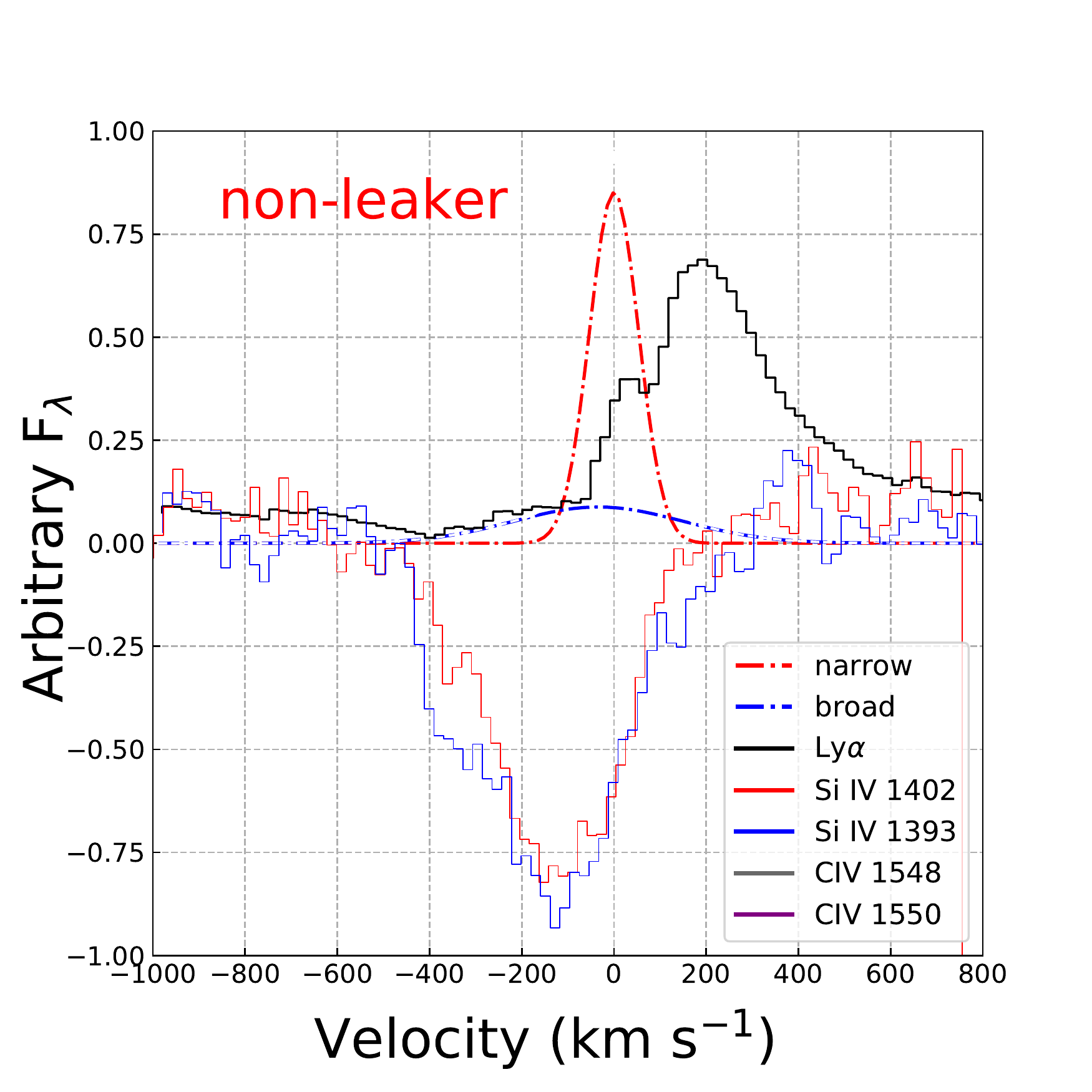}
\caption{Comparison of outflowing gas inferred from ultraviolet absorption lines and optical emission lines for the leaker (top panel) and the non-leaker (bottom panel). The velocity profiles from narrow (red curve) and broad (blue curve) components of [OIII]$\lambda$5007 emission line are plotted against low ionization absorption lines (left panel) and high ionization absorption lines (right panel). }
\label{fig:interstellar_features}
\end{figure*}

In Figure~\ref{fig:interstellar_features}, we show the different interstellar absorption features identified in the leaker and non-leaker stack, and compare them with the gas kinematics observed with the [O III] 5007 line. The resolution of the MAGE spectra provides resolved velocity profiles of the interstellar absorption lines. We normalize the spectra by the continuum level obtained from the stellar population fitting procedure as discussed in \S3.4.

For the non-leaker, the stacked MagE spectrum shows prominent absorption from low ionized gas (Si II 1260, O I 1302, C II 1334, Al III 1670).  The absorption peaks at $\sim$-150 km s$^{-1}$ relative to systemic, and extends from $\sim$+50 km s$^{-1}$ to $\sim$-500 km s$^{-1}$. The highly ionized gas, as traced by Si~IV and C~IV, shows similar gas kinematics.  

By contrast, for the leaker, the stacked MagE spectrum shows extremely weak absorption from the same four tracers of low ionized gas. A small amount of blue-shifted absorption appears at $\sim$-50 km s$^{-1}$ and $\sim-375$ km s$^{-1}$ relative to systemic.  The high ionization transitions show strong absorption at these same two velocities, and in addition, show a tail of high-velocity absorption extending out to $\sim$-750 km s$^{-1}$.

Comparing the absorption velocity profiles between the leaker and the non-leaker, while both show significant absorption from highly ionized gas, for the leaker the absorption extends  to higher velocities (-750 km s$^{-1}$) than the non-leaker (-500 km s$^{-1}$).

\section{Discussion}

\subsection{Outflows traced by UV and optical spectra}
Outflowing gas in star-forming galaxies at z$\sim$1-3 has been reported both from optical emission lines \citep[e.g.][]{Genzel2011,Newman2012,Davies2019} as well as ultraviolet absorption lines \citep[e.g.][]{Weiner2009, Steidel2010,Martin2012,Bordoloi2014,Bordoloi2016,Jones2018}. These studies have found that the outflows traced by optical emission lines typically imply a dense phase that extends out to kpc scales, and is driven by recent star formation activity. On the other hand, outflows traced by ultraviolet absorption can trace a diffuse phase that extends over larger galactic scales, accumulated over longer timescales. However, quite rare in the literature are studies of outflow, at any redshift that are traced by both ultraviolet absorption lines and optical emission lines, due to the need for high signal-to-noise data that covers both the rest-frame UV and rest-frame optical bandpasses \citep[e.g.][]{James2014,Wood2015}.

In the Sunburst Arc, we confidently detect outflowing gas from emission lines as well as ultraviolet absorption lines. As presented in \S4, the blue-shifted broad component in [OIII] 5007 implies outflowing gas with $v_{out}$ = 327$\pm$2 km s$^{-1}$ for the leaker and $v_{out}$ = 454$\pm$47 km s$^{-1}$ for the non-leaker. While we detect outflowing gas from ultraviolet absorption lines in both the leaker and non-leaker, the maximum velocity of the outflow in the leaker is considerably larger ($\sim$750 km s$^{-1}$ as measured from high ionization absorption lines) than the maximum value measured from emission lines. The non-leaker does not show this effect; the maximum velocities traced by the ultraviolet absorption lines and broad emission components are mostly consistent within  uncertainties.

This difference in maximum velocities traced by emission and absorption lines in the leaker may indicate different regimes of the outflow. As shown in Figure~\ref{fig:interstellar_features}, the absorption lines indicate outflowing material at higher velocities than implied by optical emission lines. Some of this outflowing material, traced by Si~IV, is likely in a phase with similar ionization state to that traced by [O III] emission. The lack of the outflow in emission at higher velocity may directly result from a density gradient \citep[][]{Wood2015}. Indeed in the leaker, [OII] and CIII] trace very different gas densities: the electron density from [OII] (335 cm$^{-3}$) is similar to typical z$\sim$2-3 galaxies \citep[e.g][]{Bian2010,Sanders2015,Steidel2016}, whereas the density measured from CIII] (66000 cm$^{-3}$) is significantly higher (Table~\ref{table:leaker_vs_nonleaker_ne}). 

Interestingly, the high-velocity tail of the broad emission component aligns very well with the blue-shifted Ly$\alpha$ emission. This effect appears to be true for both the leaker and non-leaker, although both broad-component and blue shifted Ly$\alpha$ emission is weaker in the non-leaker.  This might indicate that the high-ionization, high-density, and high-velocity outflow regions are the same regions of low neutral gas column density that enable the Ly$\alpha$ (and LyC) photons to escape \citep[]{Gazagnes2020}.

\subsection{Why is the leaker leaking?}
LyC escape requires both production and leakage of UV photons. LyC production is likely associated with the most vigorously star-forming region in a galaxy \citep{Ma2020}. In the Sunburst Arc, the leaker is responsible for 18$\pm$4\% of the rest-frame 1500~\AA\ flux, which can serve as a proxy for star formation rate \citep{Kennicutt1998}.  As such, the leaker is the most vigorous site of star formation within its parent galaxy and is also the dominant producer of UV photons. 

As shown in Figure~\ref{fig:stellar_age}, production of UV photons is particularly high in the leaking region, which shows the spectral hallmarks of very massive stars (P-Cygni features in NV and CIV), the kind that are most effective at generating ionizing photons. A young stellar population with stellar ages of only a few million years is the primary production of LyC photons in a galaxy. Previous studies have shown that the majority of LyC photons are produced by 1--3 Myr stellar population with some contribution from 3--10 Myr stars \citep{Ma2015, Ma2020, Kimm2017, Kim2019, Kakiichi2021}. We have shown that the portion of the Sunburst Arc that is emitting LyC is considerably younger than the rest of the galaxy, with light-weighted ages of 3.3 Myr versus 11.8 Myr. Thus, we attribute the youth of the leaker's stellar population to its prodigious production of ionizing photons.  By contrast, the older stellar population of the rest of the galaxy is unable to produce ionizing photons. 

Thus, the generation of LyC photons has a clear physical cause: young stars.  By contrast, the physical processes responsible for the escape of those photons are not well established. Two mechanisms have been proposed in the literature.  The first mechanism is mechanical feedback, in the form of SNe activity or stellar winds (which occur at younger timescales than SNe), that would clear out neutral gas surrounding the young stellar cluster, such that LyC photons can eventually escape \citep[e.g.][]{Wise2009, Heckman2011, Sharma2017, Hogarth2020}. The second mechanism is the ionization of surrounding gas from young stellar populations, thus creating a density-bounded H II region in which LyC photons penetrate through the low column density region \citep{Jaskot2013, Nakajima2014}. Both processes may act together to facilitate LyC escape.

The rest-frame UV spectra (Figure~\ref{fig:interstellar_features}) imply that the LyC photons are escaping because there is a deficit of low-ionization gas, which is present in abundance only $\sim$ 400 pc away. The presence of some neutral gas along the line of sight is evident from the weak low-ionization absorption gas. This supports some neutral gas along the line of sight and that models
of the Lyman alpha line favor a model with narrow ionized channels as per \citet{Rivera-Thorsen2017}.

\subsection{Physical scale of ionized outflow}
A key open question of modern astrophysics is how the ``feedback'' from young star clusters is able to regulate star formation over the spatial scales of a whole galaxy.  Fortunately, gravitational magnification allows the spatially resolved study of these processes in the Sunburst Arc, where a single star-forming region dominates the ionizing photon production and escapes, at a spatial resolution that is difficult to achieve in the local universe and impossible without lensing in the distant universe.

The leaking region shows clear evidence of ionized outflowing gas (Figure~\ref{fig:oiii_profile_compare}), which is considerably weaker in the non-leaking portions of the galaxy. Such a strong gas outflow is likely driven by recent star formation activity in the LyC leaking region. It is unclear whether the weak broad component seen in the non-leaker arises in the non-leaking region, or whether it is just the extension of the leaker's outflow that reaches to the non-leaking regions.  Whichever the case, it is clear that the strongly ionized outflows are spatially associated with the leaker.  The high ionization absorption lines further support this -- they are much stronger in the leaker compared to the non-leaking regions (Figure~\ref{fig:interstellar_features}).

Further examination of the individual non-leaking regions allows us to constrain the spatial extent of the outflow. Three different FIRE pointings targeting non-leaking regions probe physical regions in the source plane that are at different distances from the leaker \citep{Sharon2022}. It is clear from Figure~\ref{fig:intrinsic_outflow} that the outflows probed by [OIII]5007 emissions are strongest in the LyC region. At the distances of 400--700 pc that the non-leaking FIRE pointings probe, the intrinsic line flux in outflow reduces by $>$80 \%. Thus, we constrain the physical scale of the ionized outflow to be $<$400 pc.

\subsection{Evidence of an ionized supper bubble in the Sunburst Arc?}

A detailed picture of the physics of LyC escape requires probing the ISM structure and feedback from young stars on sub-kiloparsec spatial resolution scales. This has been challenging to achieve both theoretically and observationally. Most recently, \citet{Ma2020} studied LyC escape from 34 z$>$5 galaxies in high-resolution cosmological simulations. They found that majority of LyC escape come from very young ($<$10 Myr), vigorously star-forming regions of the galaxy. These regions may reside in a kpc-scale ionized bubble presumably created by SNe activity from 3--10 Myr stars. Low column density sight lines can be fully ionized by young stars allowing the escape of ionizing photons. In contrast, the galaxy may contain several other non-leaking regions dominated by young stars but still embedded in dense neutral clouds. Feedback from young stars in the non-leaking regions is not strong enough to clear pathways for LyC escape.

The LyC emitting region in the Sunburst Arc is the most vigorous star-forming region in the galaxy and includes the youngest stars ($\sim$ 3 Myr) in the galaxy. The stellar feedback from such a young stellar population is likely driving the ionized outflowing gas. Over the time (3--10 Myr), such ionized outflowing gas may turn into a giant ionized super bubble surrounding the star formation site. In such a scenario, the strongly localized outflows seen in emission lines of the leaker may indicate that the LyC region lies within a super bubble. 

Recently, \citet{Menacho2019} reported evidence of a kpc-scale super bubble in a local LyC emitter (Haro 11), which may have cleared channels for LyC escape in the galaxy. This seems quite analogous to the $<$400~pc bubble of ionized gas in the Sunburst Arc.  Super bubbles may be important mechanisms for enabling LyC photons to escape galaxies. An upcoming IFU study with {\it JWST} (GO: 2555, PI: T Rivera-Thorsen) will measure the physical size of the superbubble in the Sunburst Arc, and obtain a detailed picture of the gas kinematics in this remarkable ionizing lensed galaxy.

\section{Summary and conclusions}

In this paper, we present a spectroscopic study of multiple star-forming regions within the LyC emitting galaxy, the Sunburst Arc, at $z$=2.37. We obtained near-infrared spectra from Magellan/FIRE and optical spectra from Magellan/MagE. One of the primary goals of the paper is to understand the physical conditions that facilitate the production and escape of LyC photons from a galaxy. In order to achieve our goal, we generated rest-frame ultraviolet and rest-frame optical stacked spectra for two regions: those that emit ionizing photons (leaker stack) and those that do not emit ionizing photons (non-leaker stack).

 The rest-frame optical spectra reveal highly ionized gas flowing out of the Sunburst Arc. The leaker shows blue shifted broad component optical emission (FWHM = 327 $\pm$ 2 km s$^{-1}$) in addition to a narrow component (FWHM = 112 $\pm$ 1 km s$^{-1}$) component, which indicates the presence of strongly ionized outflowing gas. This outflow is strongest in the spectra of the leaker.  Indeed, most (54.5$\pm$0.3\%) of the [O III]~5007 emissions from the leaker emerges from a broad component. The stacked non-leaker also shows this broad component, but it accounts for only 26.1$\pm$7.0\% of the [O III]~5007 emissions.  Examining the spectra of the individual non-leaking components (Fig~\ref{fig:intrinsic_outflow}), the broad component decreases with increasing distance from the leaker.
 
 The high-ionization gas traced by rest-frame UV absorption lines echoes the above picture: both the leaker and the non-leaker show blue-shifted gas.  In the non-leaker this gas is centered at $-$200~km s$^{-1}$ and extends to 600~km s$^{-1}$; in the leaker the absorption has two peaks at $-$50~km s$^{-1}$ and $-$350~km s$^{-1}$, and extends all the way to $-$750~km s$^{-1}$.  Thus, the spectra of both the leaking and non-leaking regions show evidence for a high-velocity, high-ionization outflow, which in the leaker extends to higher velocities. The rest-frame UV absorption lines provide something the rest-frame optical emission lines do not:  a picture of the low ionization gas, as traced by Si II 1260, CII 1334, and Al III 1670.   The non-leaker shows absorption centered at $-$200 \kms, extending from 0 to $-$500 km/s.  In the leaker, this low-ionization gas is much weaker, with two very weak absorption features at $\sim-$50 km s$^{-1}$ and $\sim-$375 km s$^{-1}$.  
 
The rest-frame UV stack spectra show features that are characteristics of the massive stars (Figure~\ref{fig:stellar_age}). In particular, the leaker shows strong P-Cygni NV and CIV features as well as broad He II emission; all are characteristics of young massive stars. In contrast, these features are weaker in the non-leaker stack. The stellar population fitting to the rest-frame UV spectra reveals that the stellar population of leakers is considerably young ($\sim$3 Myr) than non-leaker ($\sim$12 Myr).
 
Overall, the presence of relatively young massive stars in the leaker suggests that ionizing photons are being produced within the leaking region. These ionizing photons are being carried out along the channels that are mostly cleared out by the stellar feedback as evident in the high ionization absorption lines and optical emission lines. There is evidence of little neutral gas along the line of sight as indicated by weak low ionization absorption lines (Figure~\ref{fig:interstellar_features}). This picture supports a picture in which LyC photons stream out of channels cleared through the interstellar medium. On the other hand, the absence of a younger stellar population in the non-leaking regions indicates that these are not actively producing ionizing photons. Additionally, stronger low ionization absorption features suggest the presence of substantial neutral gas along the line of sight which would absorb any ionizing photons being carried out. Together, these data support a picture in which a
young ( 3Myr) vigorous star cluster not only prodigiously generates ionizing photons, but also drives
a high-velocity outflows that clear out neutral gas along certain sight-lines, enabling the escape of ionizing photons.

The rest-frame UV stack spectra show features that are characteristic of massive stars (Figure~\ref{fig:stellar_age}). In particular, the leaker shows strong P-Cygni NV and CIV features as well as broad He II emission; all are characteristic of young massive stars. By contrast, these features are weaker in the non-leaker stack. Stellar population fitting to the rest-frame UV spectra reveals that the leaker is considerably young ($\sim$3 Myr) than the non-leaker ($\sim$12 Myr).
 
 Thus, we see a consistent picture, in which the only region of the Sunburst Arc that is known to be leaking ionizing photons --- the ``leaker'' --- is also the only region showing spectral signatures of the young massive stars that are capable of producing large numbers of ionizing photons.  The sightlines to the leaker appear to have little neutral gas along the line of sight, as indicated by weak low ionization absorption lines.  What gas is present appears to be highly ionized, as evident in the high ionization absorption lines.  The broad component of the [O III] emission suggests that stellar feedback is what has cleared out these channels. This picture supports a model for LyC escape, in which LyC photons stream out of channels cleared through the interstellar medium.
 
 By contrast, the non-leaking regions lack the signposts of a very young stellar population, have stronger low ionization absorption features that suggest substantial neutral gas along the line of sight, and have a much weaker broad velocity component of [O III].  In other words, in the non-leaking regions, we see no evidence that ionizing photons are being created, nor do we see clear paths by which ionizing photons could escape, nor do we see evidence of stellar feedback that could clear out these paths.
 The Sunburst Arc may be one of the best places to catch these feedback processes in action.



\section{Acknowledgments}.
\begin{acknowledgments}
We thank the referee for their useful comments. The author is grateful to Daniel Stark for enlightening conversation. The material is based upon work supported by NASA under award number 80GSFC21M0002. Based on observations made with the NASA/ESA {\it Hubble Space Telescope}, obtained at the Space Telescope Science  Institute, which is operated by the Association of Universities for Research in Astronomy, Inc., under NASA contract NAS 5-26555. These observations are associated with program GO-15101. Support for Program number GO-15101 was provided through a grant from the STScI under NASA contract NAS5-26555. Support for this work was provided by the National Aeronautics and Space Administration through Chandra Award Number GO8-19084X issued by the Chandra X-ray Center, which is operated by the Smithsonian Astrophysical Observatory for and on behalf of the National Aeronautics Space Administration under contract NAS8-03060. The scientific results reported in this article are based on observations made by the Chandra X-ray Observatory, and this research has made use of software provided by the Chandra X-ray Center (CXC) in the application package, CIAO.
This paper includes data gathered with the 6.5 m Magellan Telescopes located at Las Campanas Observatory, Chile. We thank the staff of Las Campanas for their dedicated service, which has made possible these observations. We thank the
telescope allocation committees of the Carnegie Observatories,
The University of Chicago, The University of Michigan, and
Harvard University, for supporting this observing program over
several years.

\end{acknowledgments}

%





\bibliography{references}{}
\bibliographystyle{aasjournal}



\end{document}